\documentclass[reprint,nofootinbib,groupedaddress,amsmath,amssymb,aps]{revtex4-1}

\usepackage{changes}
\usepackage{soul}
\usepackage[utf8]{inputenc}
\usepackage[english]{babel}
\usepackage{amsmath,amsfonts,amssymb}
\usepackage[T1]{fontenc}
\usepackage{graphicx}
\usepackage{upgreek}
\usepackage[colorlinks=true,bookmarks=false,citecolor=blue,urlcolor=blue]{hyperref} 

\usepackage{bm}

\newcommand{\cod}{CO$_2$}

\renewcommand*{\thefootnote}{\alph{footnote}}
\usepackage{enumitem}
\setlist{nolistsep}
\usepackage[toc]{appendix}

\makeatletter
\newcommand{\vast}{\bBigg@{3.5}}
\newcommand{\Vast}{\bBigg@{5}}
\makeatother

\newcommand\s[1]{\mathrm{#1}}

\usepackage{letltxmacro}
\makeatletter
\let\oldr@@t\r@@t
\def\r@@t#1#2{%
	\setbox0=\hbox{$\oldr@@t#1{#2\,}$}\dimen0=\ht0
	\advance\dimen0-0.2\ht0
	\setbox2=\hbox{\vrule height\ht0 depth -\dimen0}%
	{\box0\lower0.4pt\box2}}
\LetLtxMacro{\oldsqrt}{\sqrt}
\renewcommand*{\sqrt}[2][\ ]{\oldsqrt[#1]{#2}}
\makeatother

\newcommand\unit[3]
{
	\ifx&#1&%
	\ifx&#2&%
	\else
	10^{#2}
	\fi
	\else
	\ifx&#2&%
	#1
	\else
	#1\cdot 10^{#2}
	\fi
	\fi
	\ifx&#3&%
	\else
	\;\mathrm{#3}
	\fi	
}

\graphicspath{{./img/}}

\begin{document}
	
	\title{Giant Brillouin amplification in gas using hollow-core waveguides}
	
	\author{Fan~Yang$^{*,\dagger}$}
	\author{Flavien~Gyger$^*$}
	\author{Luc~Thévenaz}
	\affiliation{Swiss Federal Institute of Technology Lausanne (EPFL), Group for Fibre Optics, SCI-STI-LT Station 11, 1015 Lausanne, Switzerland}
	
	\def\thefootnote{*}\footnotetext{These two authors contributed equally to this work}\def\thefootnote{$\dagger$}\footnotetext{Corresponding author: fan.yang$@$epfl.ch}
	

	\begin{abstract}
		
		\noindent
		\textbf{
			Stimulated Brillouin scattering (SBS) offers among the highest nonlinear gains in solid materials and has demonstrated advanced photonics functionalities in waveguides. The large compressibility of gases suggests that SBS may gain in efficiency with respect to condensed materials. Here, by using a gas-filled hollow-core fibre at high pressure, we achieve a strong Brillouin amplification per unit length, exceeding by 6 times the gain observed in fibres with a solid silica core. This large amplification benefits from a higher molecular density and a lower acoustic attenuation at higher pressure, combined with a tight light confinement. Using this approach, we demonstrate the capability to perform large optical amplifications in hollow-core waveguides. The implementations of a low-threshold gas Brillouin fibre laser and a high-performance distributed temperature sensor, intrinsically free of strain cross-sensitivity, illustrate the large perspectives for hollow-core fibres, paving the way to their integration into lasing, sensing and signal processing.} 
		
	\end{abstract}
	
	\maketitle
	
	\begin{figure*}[htbp]
		\centering{
			\includegraphics[width = 1 \linewidth]{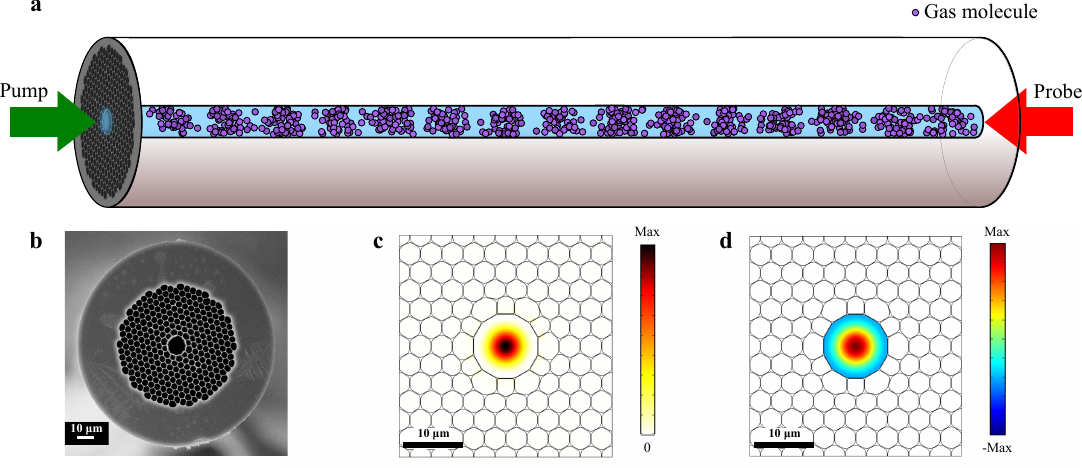}
		}
		\caption{\textbf{Principle of the generation of stimulated Brillouin scattering (SBS) in gas-filled hollow-core fibres (HCFs).}
			(\textbf{a}) Conceptual view of the interacting waves in the SBS process: the pump and probe optical waves are separately launched into each HCF end and counterpropagate. Their interference creates an intensity beat pattern that slowly moves along the fibre owing to their slight frequency difference. The electrostrictive force on the gas molecules causes periodic density fluctuations thanks to the gas compressibility. This periodic density distribution is experienced as a moving refractive index grating by the optical waves that are consequently coupled. The process turns resonant when the beat pattern moves exactly at the sound velocity in the medium, which is realised for a well-defined frequency difference between the optical waves. In this case a strong unidirectional energy transfer is observed from pump to probe and the probe is amplified.
			(\textbf{b}) Scanning electron microscope image of the HCF used in this work. The fibre presents a core diameter of 10.9 ${\rm \upmu m}$ and a cladding diameter of 120 ${\rm \upmu m}$.
			Spatial distributions of (\textbf{c}) the intensity of the fundamental optical mode and (\textbf{d}) the acoustic amplitude of the first excited radial acoustic mode in the HCF. The SBS efficiency is scaled by the overlap integral between these 2 distributions, which is highest for the 2 presented modes.
			\label{fig_Introduction}
		}
	\end{figure*}

	\noindent
	Stimulated Brillouin scattering (SBS) is a third-order optical nonlinear effect that manifests itself in a coherent light-sound coupling \cite{eggleton_brillouin_2019,safavi-naeini_controlling_2019,wiederhecker_brillouin_2019,kobyakov2010stimulated}. It is usually the strongest nonlinear effect in amorphous materials \cite{eggleton_brillouin_2019} and has been observed in many platforms, such as optical fibres \cite{ippen_stimulated_1972,dainese_stimulated_2006,kang_tightly_2009}, whispering gallery mode resonators \cite{grudinin_brillouin_2009,tomes_photonic_2009,lee_chemically_2012}, integrated waveguides \cite{pant_-chip_2011,shin_control_2015,van_laer_interaction_2015,yang_bridging_2018,gundavarapu_sub-hertz_2019} and various fluids \cite{hagenlocker_stimulated_1965,she_stimulated_1983,manteghi_spectral_2011,mountain_spectral_1966,dahan_droplet_2016,giorgini_stimulated_2018}.  
	
	Gas turns out to be an attractive medium for nonlinear optics because, unlike condensed matter, it is not subject to optical damage at high intensities (with the notable exception of photochemical dissociation for some molecular gases). It shows a pressure-dependent nonlinearity and group-velocity dispersion, as well as a potentially wider transparency window from the vacuum ultraviolet to the mid-infrared region \cite{russell_hollow-core_2014,travers_ultrafast_2011}. Various nonlinear effects have been demonstrated in gases, including supercontinuum generation \cite{corkum_supercontinuum_1986}, high-harmonic generation \cite{popmintchev_attosecond_2010}, filamentation \cite{berge_ultrashort_2007}, Raman scattering and Brillouin scattering \cite{hagenlocker_stimulated_1965}.
	
	So far, backward SBS in gases has been exclusively observed using free space interactions \cite{hagenlocker_stimulated_1965,she_stimulated_1983,manteghi_spectral_2011}. The scattering efficiency remains limited owing to the weak light confinement over a sizeable interaction length, so that high-power laser pulses (megawatt peak power) are needed to eventually observe a moderate SBS signal \cite{hagenlocker_stimulated_1965,she_stimulated_1983,manteghi_spectral_2011}. 
	
	Hollow-core fibres (HCFs) including hollow-core photonic bandgap fibres, Kagome-style hollow-core fibres and single-ring antiresonant fibres, show low-loss and diffraction-less optical transmission (state of the art loss: 0.65 dB/km \cite{bradley_antiresonant_2019}). They are therefore the ideal candidate to drastically increase the light-sound interaction length in a gaseous medium and thereby achieve an efficient coupling between interacting waves \cite{russell_hollow-core_2014,travers_ultrafast_2011,dudley_ten_2009}. In free space, a Gaussian beam converging to a spot of diameter $2a$ results in a nonlinear gain proportional to $L_{\rm {eff}}/\pi a^2$, where $L_{\rm {eff}}$ is the effective interaction length. $L_{\rm {eff}}$ is given by twice the Rayleigh length $z_{\rm R}=\pi a^2/\lambda$, where $\lambda$ is the laser wavelength in vacuum. As a consequence, the nonlinear gain in bulk gases is proportional to $2z_{\rm R}/\pi a^2 = 2/\lambda$, irrespective of the beam convergence.
	
	On the other hand, in gas-filled HCFs, $L_{\rm {eff}}$ is solely function of the fibre length and attenuation and can actually reach a few hundred metres. Additionally, the mode field diameter is typically $\sim 10~ \rm {\upmu m}$, so that the nonlinear gain in gas-filled HCFs may turn six orders of magnitude larger than in unconfined gases (i.e. in free space) in the near-infrared region. Incidentally, the threshold for nonlinear effects, such as stimulated Raman scattering \cite{benabid_stimulated_2002} and Raman frequency comb generation \cite{couny_generation_2007}, has been proved to be several orders of magnitude lower than using free-space optics. 
	
	It has to be nevertheless mentioned that an opto-acoustic interaction, in this case forward SBS, has been recently demonstrated in HCF filled with air at atmospheric pressure \cite{renninger_guided-wave_2016}. In that work, the peak gain coefficient reached ${4 \times 10^{-14} ~{\rm m/W}}$, which remains $1000$ times smaller than the peak backward Brillouin gain coefficient in fibres with a solid silica core (SMF). This gain is clamped down as a consequence of the non-uniformity of the core diameter \cite{jarschel_fiber_2018}, since it causes a inhomogeneous gain linewidth broadening. Globally, no strong light-sound interaction in gaseous media has been reported so far.
	
	Here, we report a considerable optical amplification by using backward SBS in a gas-filled HCF. We achieve 0.32 dB of signal amplification per mW of pump power inside a 50 m long HCF filled with carbon dioxide ($\rm {CO_2}$) at a pressure of 41 bar. This large gain results from two causes: firstly, the peak Brillouin gain coefficient shows a quadratic dependence on the gas pressure, in contrast with stimulated Raman scattering \cite{russell_hollow-core_2014} and Kerr nonlinearity \cite{travers_ultrafast_2011} in which the nonlinear coefficients are typically proportional to the gas pressure. The Brillouin gain can therefore be drastically enhanced in the backward SBS configuration by raising the gas pressure. Secondly, the simultaneous tight confinement of light and gas in a HCF offers altogether an ultra-long interaction length and a small effective beam cross-section.
	Using this platform, we demonstrate two original and specific implementations: a low-threshold (33 mW) continuous-wave single-frequency laser in a HCF that can in principle operate at any wavelength, as well as a distributed temperature sensor of unprecedented performance showing zero strain cross-sensitivity, thereby breaking a 30-year physics barrier since Brillouin fibre sensing was first proposed. \\

	\noindent
	{\large{\textbf{Results}}} \\
	\noindent
	{\large{\textbf{Theoretical analysis of the Brillouin gain}}}
	
	\noindent
	Stimulated Brillouin scattering in a gas-filled HCF is a process in which a pump and probe optical waves with a slightly different frequency counter-propagate along a HCF and their interference produces a longitudinally moving fringe pattern. When a strict phase matching condition is met, dictated by the relative velocities of light and sound in the medium, the fringe pattern gives rise, via electrostriction, to a travelling longitudinal acoustic wave in the gas, as illustrated in Fig. \ref{fig_Introduction}(a). In turn, this wave periodically modifies the medium optical density, inducing a Bragg-type coupling between pump and probe. We assume the frequency of the pump light to be higher than that of the probe. In this case, the probe is amplified by the pump when their frequency difference matches the Brillouin frequency shift (i.e. the frequency for perfect phase matching), given by $\nu_{\rm B}=2n_{\rm eff}v_{\rm a}/\lambda$, where $n_{\rm eff}$ is the effective refractive index of the optical mode, $v_{\rm a}$ is the gas acoustic velocity and $\lambda$ is the pump wavelength. Note that Brillouin amplification in gas can be implemented in any hollow-core waveguide, including microstructured fibres, capillary fibres and metal-coated waveguides, as well as slot waveguides. For this first demonstration, we opted for a 50 m long commercial HC-1550-02 HCF from NKT Photonics, since it shows a relatively small optical effective mode area (51 $\upmu$m$^2$), leading to a higher Brillouin coefficient. A scanning electron microscope image of its cross-section is shown in Fig. \ref{fig_Introduction}(b). This fibre guides light inside its hollow core by virtue of the photonic bandgap in the periodically patterned cladding region. 
	
	During the SBS interaction, the probe wave is amplified with a peak gain given by \cite{kobyakov2010stimulated}:

	\begin{equation}
		\label{eq_gain}
		{
			g_{\rm B} = \frac{\gamma^2_{\rm e}\omega^2}{\rho n v_{\rm a} c^3 \Gamma_{\rm B}  A^{\rm ao}_{\rm eff}}},
	\end{equation}
	where $\gamma_{\rm e}$ is the electrostrictive constant in the gas medium, $\omega$ is the light angular frequency, $\rho$ is the gas density, $n$ is the gas refractive index, $c$ is the light velocity in vacuum, $\Gamma_{\rm B}/2\pi$ is the gain spectrum linewidth (full width at half maximum (FWHM)), directly related to the acoustic attenuation, and $A^{\rm ao}_{\rm eff}$ is the acousto-optic overlap effective area, as defined and calculated in Supplementary Section S4. Figures \ref{fig_Introduction}(c) and (d) show the spatial distributions of the fundamental optical and first excited radial acoustic eigenmodes, respectively (obtained by a finite element method using COMSOL Multiphysics). $A^{\rm ao}_{\rm eff}$ is calculated from the overlap integral between these 2 distributions. In this work, we consider only the first excited radial acoustic mode, since the Brillouin interaction involving higher acoustic modes is more than 2 orders of magnitude smaller.
	
	The acoustic velocity is given by $v_{\rm a}=1/\sqrt{\beta_{\rm s}\cdot\rho}$, independent of pressure for an ideal gas, where $\beta_{\rm s}$ is the adiabatic compressibility. As the compressibility coefficient of gases (e.g. at 40 bar) is four orders of magnitude larger than that of solid materials (e.g. fused silica) and the density of gases (at 40 bar) is some 30 times smaller than that of a condensed material, the acoustic velocity is about 20 times smaller. This leads to a 20-fold increase in gain, as shown by Eq. (\ref{eq_gain}), as well as a Brillouin frequency shift $\nu_{\rm B}$ lying in the sub-gigahertz range (e.g. $\sim 320$ MHz for \cod\ at a pump wavelength of 1.55 $\rm {\upmu m}$).

	We shall now discuss the two key-parameters contributing to the quadratic dependence of the Brillouin gain on gas pressure: the electrostrictive constant, $\gamma_e$, and the acoustic attenuation coefficient, $\Gamma_\s{B}$.
	
	The \textit{electrostrictive constant} is defined as the normalised rate of change of the relative permittivity $\epsilon_r$ in response to a change in the density $\rho$ \cite{boyd_nonlinear_2008}: 
	\begin{equation}
		\gamma_{\rm e} = \rho\frac{\partial\epsilon_{\rm r}}{\partial\rho} = \rho\frac{\partial\chi}{\partial\rho},
	\end{equation}
	where $\chi$ is the electric susceptibility. At the pressure ranges considered in this manuscript, the electric susceptibility of \cod\ shows a linear dependence on density: $\chi = A\rho$, where $A\approx 5\times10^{-4}~\s{m^3/kg}$ for \cod. As a result, the electrostrictive constant:
	\begin{equation}
		\gamma_{\rm e} = A\rho
		\label{eq_electrostriction}
	\end{equation}
	is directly proportional to the density, hence to the pressure in the ideal gas approximation.

	The \textit{acoustic attenuation} of a sound wave in \cod\ at megahertz frequencies arises mainly from two dissipative phenomena: viscous forces and thermal conductivity \cite{bhatia1985ultrasonic}. Thermal conductivity and viscosity have very similar origins: they arise from energy/momentum diffusion, respectively, driven by thermal motion of the gas molecules which compensates for any temperature/velocity gradient. Hence, their strengths are proportional to the temperature/velocity gradient, respectively. In order to intuitively grasp the pressure dependence in these two processes, let us consider an acoustic plane wave propagating along the $x$-axis in a gas that is globally at rest. Such a wave consists of similar periodic variations of density, pressure, velocity and temperature. In particular, we express the velocity of the gas volume elements as: $u(x,t) = u_0\cos{(qx-\Omega t)}$, where $u_0$ is the velocity amplitude, $\Omega$ and $q$ are the wave angular frequency and wavenumber, respectively (for backward SBS, $q\approx2n\omega/c$). The intensity of the acoustic plane wave is expressed as:
	\begin{equation}
		I_{\rm {ac}} = \frac{1}{2}\rho\cdot v_{\rm a} \cdot u_0^2. 
	\end{equation}
	Within the ideal gas model, it can be shown that the temperature oscillations of the acoustic wave are given by \cite{bhatia1985ultrasonic}:
	\begin{equation}
		T(x,t) = \frac{T_0}{v_a}\left(\gamma-1\right)u_0\cos{(qx-\Omega t)},
	\end{equation}
	where $T_0$ is the average ambient temperature and $\gamma$ is the adiabatic index, namely the ratio of specific heats at constant pressure and volume, respectively. Thus, for an intensity kept constant, both gas velocity and temperature periodic variations (and hence, their gradient) decrease for an increasing density $\rho$. As a consequence, both thermal diffusion and viscosity forces are in proportion equally reduced. It can be more formally shown that the dissipated energy caused by each process is proportional to $u_0^2$ and the acoustic attenuation coefficient is therefore proportional to $1/\rho$, expressed as \cite{landau1987fluid}:
	\begin{equation}
		\label{eq_linewidth}
		\Gamma_{\rm B} = \frac{q^2}{\rho}\left[\frac{4}{3}\eta_{\rm s}+\eta_{\rm b}+\frac{\kappa}{C_{\rm P}}(\gamma-1)\right],
	\end{equation}
	where $\eta_{\rm s}$ and $\eta_{\rm b}$ are the shear and bulk viscosity respectively, $\kappa$ is the thermal conductivity and $C_{\rm P}$ is the specific heat at constant pressure. Inserting Eqs. (\ref{eq_electrostriction}) and (\ref{eq_linewidth}) into Eq. (\ref{eq_gain}), it ends up that the Brillouin gain increases quadratically with the density, thus the pressure in the ideal gas approximation. \\

	\begin{figure*}[htbp]
		\centering{
			\includegraphics[width = 0.8 \linewidth]{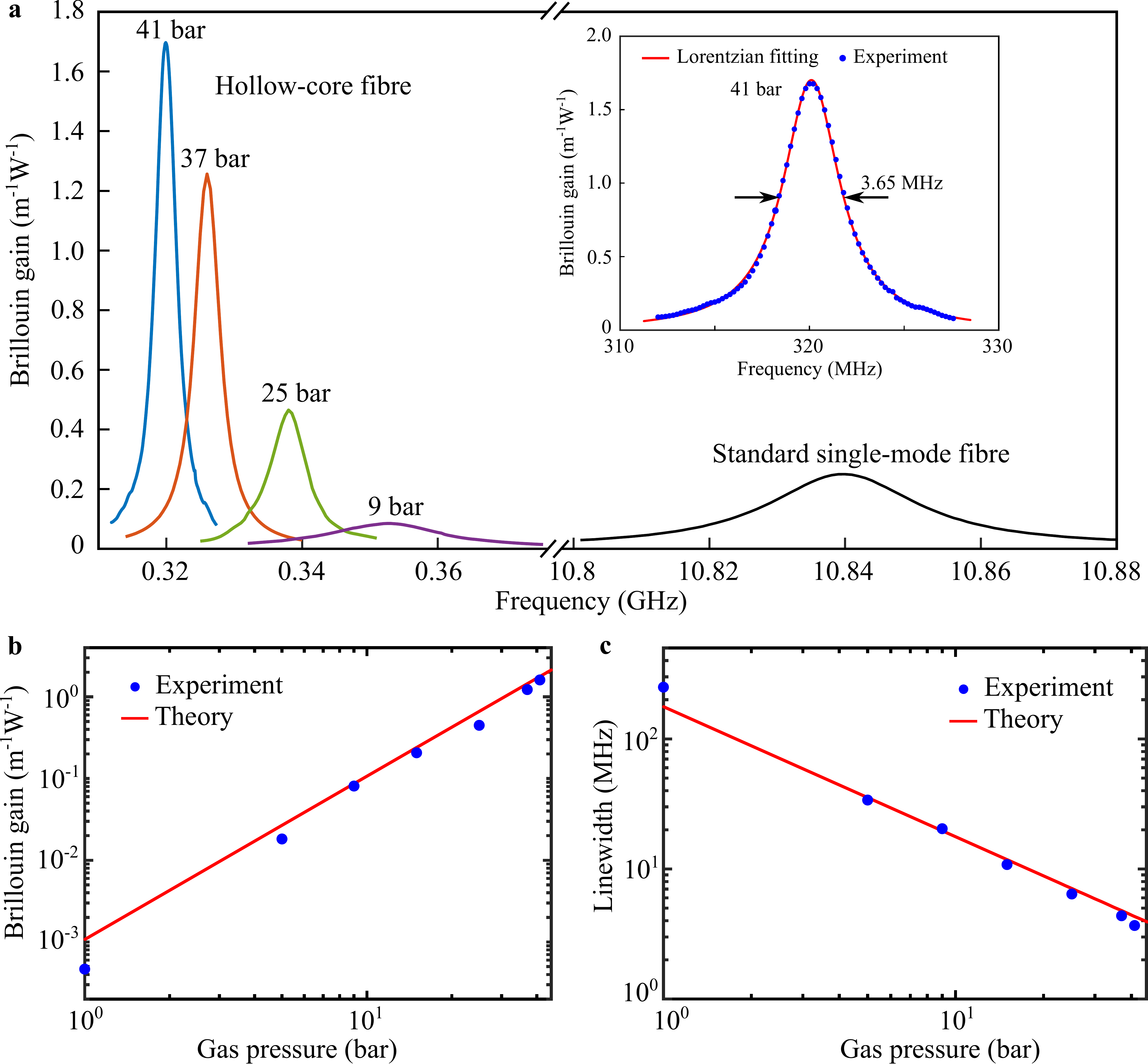}
		}
		\caption{\textbf{Experimental gains in gas by stimulated Brillouin scattering.}
			(\textbf{a}) Measured Brillouin gain spectra in a hollow-core fibre filled with ${\rm CO_2}$ at different pressures. The Brillouin spectrum of a solid silica-core single-mode fibre is also shown for comparison. Note the horizontal scale is discontinued, but its intervals are kept constant. Inset: magnified view of the 41 bar-filled HCF gain spectrum, showing measured datapoints and Lorentzian fitting.
			(\textbf{b}) Measured peak Brillouin gain per unit fibre length and unit pump power and (\textbf{c}) full width at half maximum (FWHM) linewidth of stimulated Brillouin scattering in ${\rm CO_2}$-filled HCF as a function of the gas pressure. The theoretical lines in (\textbf{b}) and (\textbf{c}) are calculated from Eq. (\ref{eq_gain}) and Eq. (\ref{eq_linewidth}), respectively. All the parameters used for these calculations are listed in Supplementary Section S4.
			\label{fig_GainMeasurement}
		}
	\end{figure*}

	\noindent
	{\large{\textbf{SBS gain coefficient measurement}}}

	\begin{figure*}[htbp]
		\centering{
			\includegraphics[width = 0.9 \linewidth]{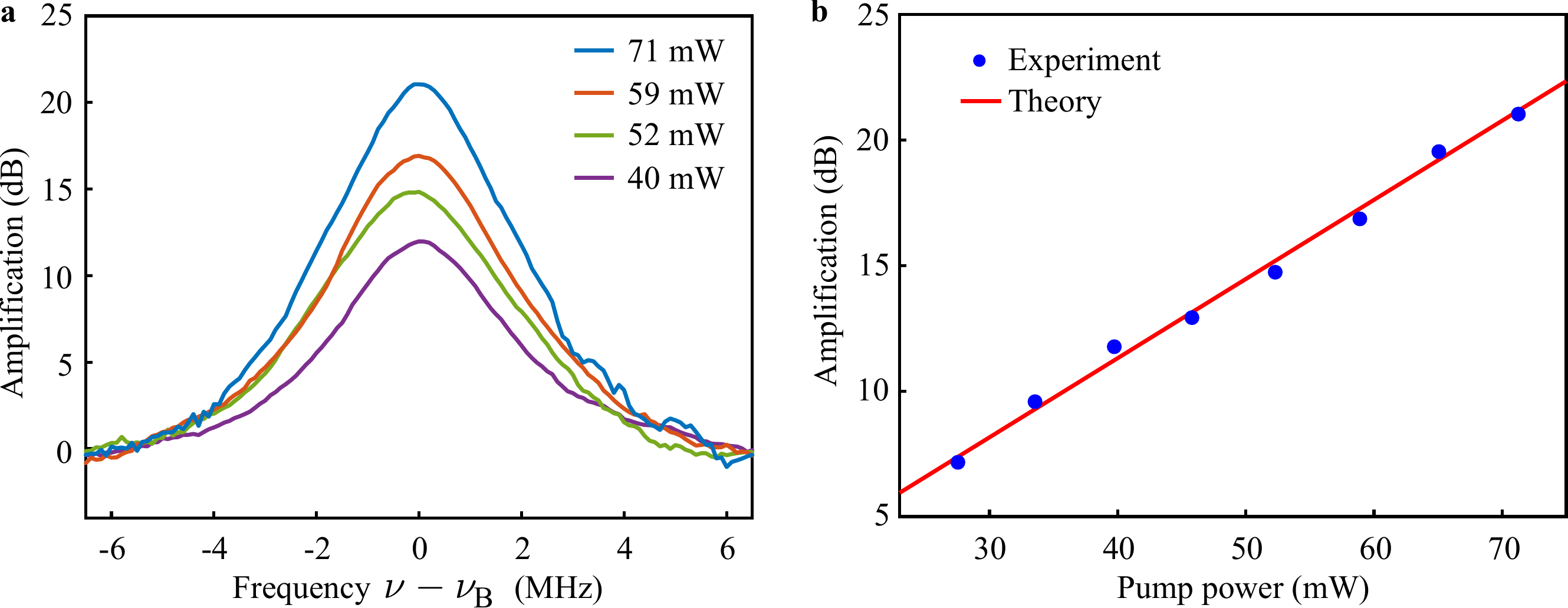}
		}
		\caption{\textbf{Amplification of the probe wave as a function of the pump power in the hollow-core fibre (HCF).}
			(\textbf{a}) Probe amplification spectra for pump powers in the HCF of 40, 52, 59 and 71 mW, respectively. Noise from spontaneous scattering is visible in the wings of the gain spectra for high pump powers.
			(\textbf{b}) Peak probe amplification in logarithmic scale as a function of pump power inside the HCF in linear scale. The dots are experimental results and the line is theoretically calculated from the equation: $G=\exp{\left(g_{\rm B}P_{\rm pump}L_{\rm eff}-\alpha L\right)}$, where $G$ is the gain, $g_{\rm B}$ is the Brillouin linear gain calculated from Eq. (\ref{eq_gain}), $P_{\rm pump}$ is the pump power at the input to the HCF (inside the HCF), $L_{\rm eff}$ is the effective length of the HCF, $\alpha$ is the fibre attenuation and $L$ is the physical fibre length. All the parameters used for the theoretical calculation are listed in Supplementary Section S2.
			\label{fig_Amplification}
		}
	\end{figure*}

	\noindent
	The \cod\ Brillouin gain at atmospheric pressure is about $10^{-13}$ m/W. In order to measure such a small gain, a lock-in detection technique has been the preferred approach to separate the gain signal from the background noise and spurious signals. The experimental set-up is detailed in Supplementary Section S1.1. The loss of our HCF filled with 1-bar \cod\ is 16 dB/km at our working vacuum wavelength of 1.55 ${\rm \upmu m}$. After gas pressurisation at 41 bar, an additional 0.5 dB loss was measured, caused by the pressure-broadened molecular absorption lines of \cod. All the experiments were performed at an environmental temperature of 24$\pm 1^{\circ}$C.
	
	Figure \ref{fig_GainMeasurement}(a) shows the measured backward SBS gain spectra of the HCF filled with \cod\ at different pressures and, for comparison purpose, of a standard SMF having a solid silica core of very similar diameter. The detailed analysis of the system response is presented in Supplementary Section S1. It can be observed that the Brillouin gain exceeds that of the standard SMF for pressures above $\sim 20$ bar. Remarkably, when the pressure reaches 41 bar, the measured Brillouin gain coefficient is 1.68 ${\rm m^{-1}W^{-1}}$, which turns out to be 6 times higher than in a standard SMF and 20 times higher than the largest Raman gain achievable in gas-filled HCFs at a wavelength of 1.55 $\rm {\upmu m}$ (i.e. at pressures above 10 bar, the peak Raman gain from the hydrogen Q(1) transition saturates to 0.08 ${\rm m^{-1}W^{-1}}$, see Supplementary Section S5 for a comparative Raman gain analysis). Since the acoustic velocity is of the order of hundreds of meters per second in gaseous media, the Brillouin frequency shift lies in the sub-GHz range, in contrast with the ~11 GHz in silica. Note that the acoustic velocity derived from the Brillouin frequency shift shown in Fig. \ref{fig_GainMeasurement}(a) decreases with rising pressure (see Supplementary Section S6 for the detailed analysis), caused by a moderate departing from the ideal gas model.
	
	For a 41-bar pressure, the $\rm {CO_2}$ Brillouin linewidth is measured to be 3.65 MHz using a Lorentzian fitting over the experimental spectrum, as shown in Fig. \ref{fig_GainMeasurement}(a) inset. This value is 10 times smaller than in a SMF, which means that the acoustic lifetime in the gas is 10 times longer than in a silica core. Figures \ref{fig_GainMeasurement}(b) and \ref{fig_GainMeasurement}(c) show the Brillouin gain and linewidth measured as a function of pressure. The measured gain coefficients and linewidths match very well with the theoretical model given by Eqs. (\ref{eq_gain}) and (\ref{eq_linewidth}): the gain is proportional to the square of the pressure, while the linewidth is inversely proportional to the pressure. During this study the maximum gain has been obtained by pressurising \cod\ at 41 bar in our 50 m long HCF, so that this configuration has been preferably used in the subsequent experiments.
	
	This HCF propagates several guided optical modes and is therefore not strictly single-moded. The light launching conditions make the fundamental mode to be much preferably populated, so that the conditions are close to a single-mode operation. This is indirectly evidenced by the symmetry and the absence of side peaks in the Brillouin gain spectra. If ever some light propagates in the higher order modes, it would lead to an underestimation of the gain value at worst. \\

	\begin{figure*}[htbp]
		\centering{
			\includegraphics[width = 1 \linewidth]{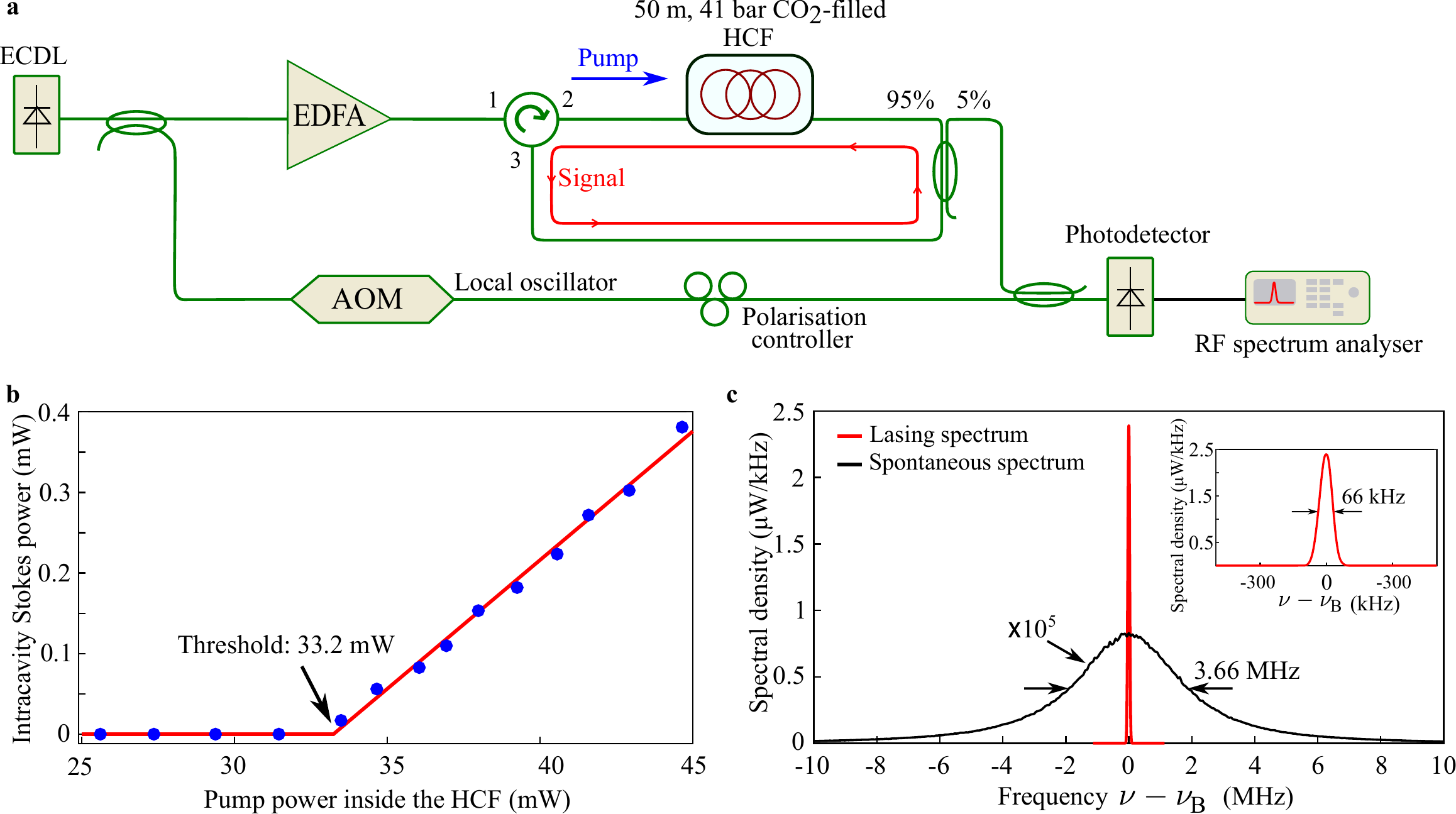}
		}
		\caption{\textbf{Gas Brillouin lasing.}
			(\textbf{a}) Experimental set-up. The continuous-wave output of an external-cavity diode laser (ECDL, Toptica CTL1550, linewidth = 10 kHz) is split: one branch is amplified by an erbium-doped fibre amplifier (EDFA) and used to pump the laser; the other branch is frequency-shifted by an acousto-optic modulator (AOM) and combined with the cavity Stokes emission for heterodyne mixing. The radio-frequency (RF) spectrum analyser resolution and video bandwidth was set to 62 kHz and 160 kHz, respectively.  
			(\textbf{b}) Intracavity Stokes optical power as a function of pump power inside the HCF. 
			(\textbf{c}) Heterodyne electrical spectrum of the Brillouin lasing emission when the pump power is 44.6 mW (above threshold). Inset: zoomed-in view of the lasing spectrum. For comparison, the heterodyne beating spectrum of the amplified spontaneous Brillouin-scattered Stokes light is also shown (highly magnified), obtained using a single-pass set-up (opened cavity) when the pump power is 12 mW inside the HCF. The lasing beating linewidth (66 kHz) is much narrower than the spontaneous spectrum linewidth (3.66 MHz), though certainly not representing the real much narrower laser linewidth. 
			\label{fig_Brillouinlasing}
		}
	\end{figure*}

	\noindent
	{\large{\textbf{Signal amplification}}}
	
	\noindent
	The evident primary application taking advantage of this large gain value is to raise the challenge of optical amplification within a HCF. To this end, we measure the amplification of a $-$34 dBm input probe beam as a function of pump power. The injected probe power is more than 40 dB smaller than the pump power, hence satisfying the small-signal amplification condition (i.e. absence of pump depletion) at least up to 30 dB amplification. The detailed set-up is shown in Supplementary Section S2.
	
	By scanning the pump-signal detuning frequency, the measured amplification spectra for different pump powers are shown in Fig. \ref{fig_Amplification}(a), while the logarithmic peak amplification value in the HCF as a function of the pump power is plotted in Fig. \ref{fig_Amplification}(b). The red line in Fig. \ref{fig_Amplification}(b) is calculated from the equation: $G=\exp{\left(g_{\rm B}P_{\rm pump}L_{\rm eff}-\alpha L\right)}$ using actual gas and fibre parameters. It shows a slope of 0.32 dB/mW, indicative of the amplification efficiency normalised to the pump power. The experimental results are in perfect agreement with the theoretical prediction. A record 53 dB amplification has been observed for a signal input power below -49 dBm and a pump power of 200 mW inside the HCF (pump-depleted regime).
	
	A straightforward estimation shows that the amplification coefficient could be enhanced up to 1.2 dB/mW by extending the effective length to 160 m using the same 41-bar \cod-filled hollow-core photonic bandgap fibre ($\sim 10~\rm {\upmu m}$ core diameter). This estimation takes into account the 26 dB/km optical loss (16 dB/km of fibre loss and 10 dB/km of molecular absorption loss in 41-bar \cod).
	
	The pros and cons of this Brillouin amplification in gas do not fundamentally differ from those extensively reported in silica-core fibres \cite{olsson_characteristics_1987}: very narrow-band efficient amplification that can be spectrally enlarged by broadening the pump spectrum through modulation, at the expense of a lowered efficiency, and poor noise figure that has still to be quantified in the case of amplification in a gas.\\

	\begin{figure*}[htbp]
		\centering{
			\includegraphics[width = 0.9 \linewidth]{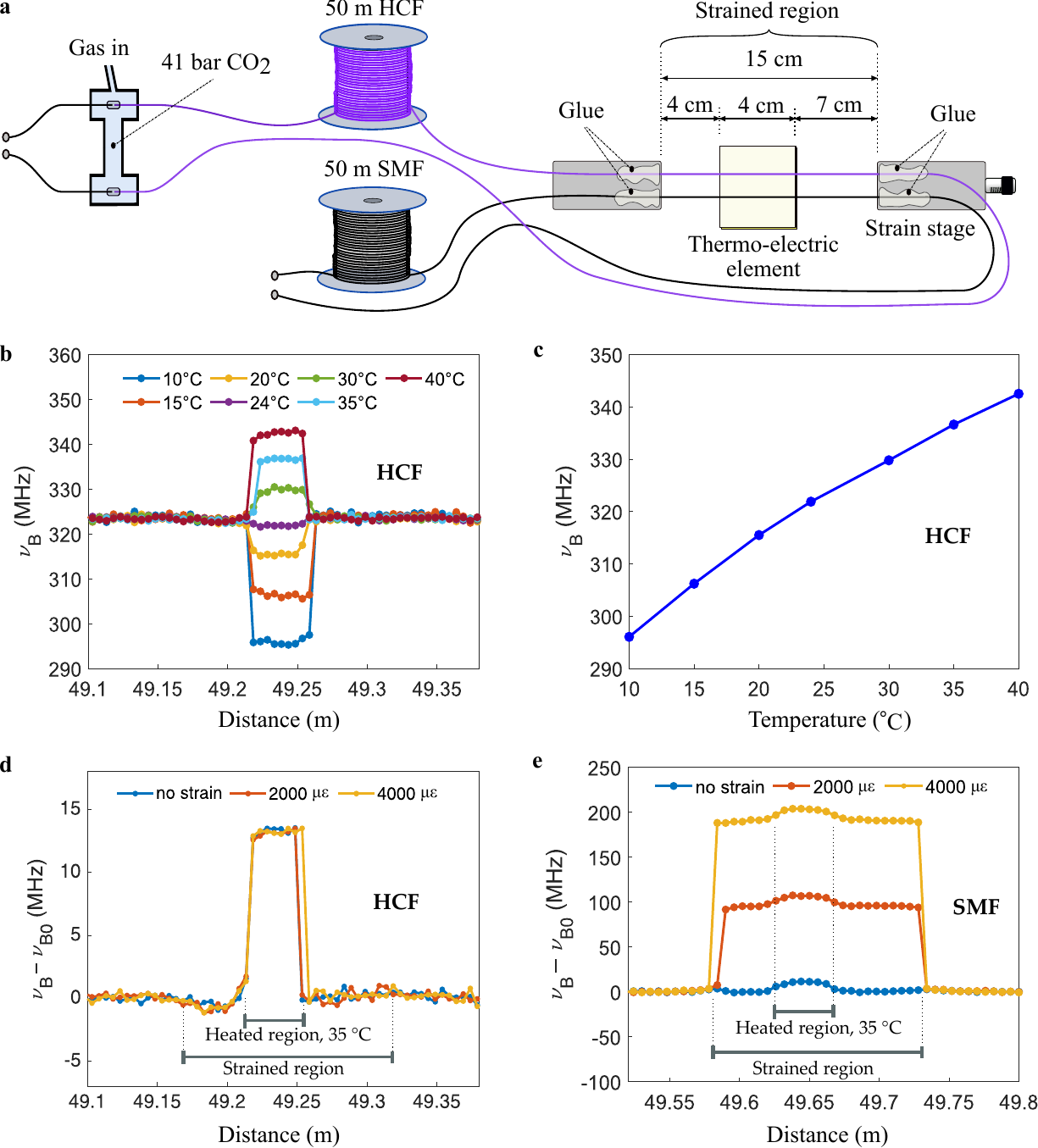}
		}
		\caption{\textbf{Distributed temperature sensing with no strain cross-sensitivity.}
			(\textbf{a}) Sensing test bench. Preset strains and temperatures can be applied on the same segment of a gas-filled HCF and a solid silica-core single-mode fibre, so that each fibre can be alternatively measured under the same conditions for a fully demonstrative comparison.
			(\textbf{b}) Longitudinal distribution of the Brillouin frequency shift in the HCF in the region of the thermo-electric element, for several preset temperatures and no applied strain. 
			(\textbf{c}) Brillouin frequency shift in the HCF as a function of the preset temperature, showing a quasi-linear dependence with a typical slope of $\sim$ 1.2 MHz/$^{\circ}$C. 
			(\textbf{d}) Longitudinal distribution of the Brillouin frequency shift in the HCF in the region of the strained section, at a preset temperature of 35$^{\circ}$C and under different applied strains, demonstrating a non-observable cross-sensitivity. 
			(\textbf{e}) Same graph as (\textbf{d}) in the SMF under identical experimental conditions, showing the presence of a massive cross-sensitivity (larger vertical scale). $\nu_{B0}$ is the Brillouin frequency shift of the fibre in ambient conditions.
			\label{fig_Distributed}
		}
	\end{figure*}

	\noindent
	{\large{\textbf{Gas Brillouin laser}}}
	
	\noindent
	This platform can be straightforwardly turned into a gas Brillouin laser by looping the 50 m long gas-filled HCF, so as to form a fibre ring cavity. Figure \ref{fig_Brillouinlasing}(a) shows the detailed experimental implementation. The pump light is launched into the cavity through a circulator and, after one revolution, is stopped by that same circulator. Since the pump is not circulating in the cavity, it must not be stringently resonant. In contrast, the Brillouin-amplified light freely circulates in the ring cavity. The total cavity length is 55 m, made of the 50-m HCF connected to 5 m of diverse SMF patchcords (circulator and coupler pigtails), so that the free spectral range of the cavity is 5 MHz, roughly equivalent to the Brillouin gain linewidth of \cod\ at 41 bar.
	
	Figure \ref{fig_Brillouinlasing}(b) shows the Stokes power as a function of the pump power evaluated inside the HCF. Brillouin lasing is turned on when the net Brillouin gain exceeds the round trip loss of the fibre cavity (i.e. 9 dB HCF insertion loss and 2 dB circulator and coupler loss, 11 dB altogether). The measured threshold is 33.2 mW, in good agreement with the theoretically estimated threshold of 34.9 mW. Considering the coupling losses, far from being optimised, this corresponds to a net pump power of 105 mW.
	
	Figure \ref{fig_Brillouinlasing}(c) shows the heterodyne electrical beating spectra between the frequency-shifted pump laser (as a local oscillator) and the amplified spontaneous Brillouin scattering (50 m long HCF filled with 41-bar \cod, single-pass backscattering through a non-looping cavity) when the pump power inside the HCF is 12 mW, as well as the beat note between the local oscillator and the Brillouin laser emission after closing the cavity, when the pump power inside the HCF is 44.6 mW. The beating spectrum linewidth (FWHM) between the Stokes signal and the local oscillator above threshold is measured to be 66 kHz, which is much narrower than the spontaneous spectrum (3.66 MHz). Since our lasing cavity is neither locked nor isolated from the environment, mode hopping constantly occurs during laser emission. In order to snapshot the heterodyne spectrum during lasing, the resolution bandwidth of the radio-frequency (RF) spectrum analyser is set to 62 kHz to promptly scan a several mega-hertz frequency range. The measured beating linewidth is therefore dominated by this resolution and does not represent the real lasing linewidth, expected to be ultra-narrow. 
	
	It should be mentioned that suppression of the mode hopping is possible using reported techniques, for instance by locking the pump-Stokes detuning frequency to a local radio-frequency oscillator \cite{danion_mode-hopping_2016}. \\

	\noindent
	{\large{\textbf{Distributed temperature sensing}}}
	
	\noindent
	Temperature/strain cross-sensitivity is currently a crucial issue in all Brillouin-based sensing systems, because the acoustic velocity in a solid is indistinctly sensitive to both these quantities that will identically impact the Brillouin frequency shift. Many methods were proposed to solve this issue by measuring two parameters showing distinct responses to temperature and strain \cite{motil_state_2016}. However, no solution, solely based on Brillouin scattering and showing intrinsic strain insensitivity, has been reported so far. The absence of cross-sensitivity is an essential quality of a sensing system. 
	
	Raman distributed sensing in silica fibre is known to show no strain cross-sensitivity. However, due to the weak response of spontaneous Raman scattering, the spatial resolution remains limited to $\sim 1~\rm m$ \cite{bolognini_raman-based_2013} and the distance range to some 30 km \cite{hartog_introduction_2018} for a reasonable integration time, far from competing with the performance of a Brillouin-based sensor.
	
	Here, we demonstrate an intrinsically strain-insensitive system based on SBS in gas-filled HCFs. In our system, optical signals keep confined into a gaseous medium, so that this configuration offers unique properties absent in solid waveguides. The absence of plasticity and stiffness of the gaseous medium leads to an insensitivity to any strain applied to its surrounding walls. Here, we take advantage of this specific properties, combined with the large Brillouin gain and its narrow linewidth to perform high performance strain-insensitive temperature measurements.
	
	We used the same 50 m long HCF filled with 41-bar ${\rm CO_2}$ and set up a phase-modulated Brillouin optical correlation-domain analysis system \cite{denisov2016going} (the detailed set-up is shown in Supplementary Section S3). The Brillouin dynamic grating position is scanned all along the fibre to measure the local Brillouin gain spectrum at each position. A strain applied on the HCF turns out to have a negligible impact on both the gas pressure and the effective optical refractive index and therefore presents no observable effect on the Brillouin frequency shift (see Supplementary Section S7 for the detailed simulation and analysis). In contrast, a change in temperature significantly modifies the acoustic velocity \cite{estrada1998speed} and hence shifts the Brillouin frequency.
	
	Our sensing system is depicted in Fig. \ref{fig_Distributed}(a). For the sake of comparison, identical lengths of HCF and solid-core SMF (ITU G.652) are jointly placed on a test bench consisting of a 4 cm thermo-electric (Peltier) element positioned in the middle of a 15 cm variable strain stage. This enables us to simultaneously apply strain and temperature changes over the same segment and identically for the two fibre types. The small thermo-electric element size turns out to be also decisive to validate the system's high spatial resolution. For each fibre, the spatial resolution was set to the highest value to secure a signal-to-noise ratio in excess of 10 at the peak gain value. The resulting spatial resolution was 1.28 cm and 2.32 cm for the HCF and SMF, respectively (calculated as the inverse of the bit duration), reflecting the contrasted difference in gain in the two media, despite a higher total loss through the HCF.
	Measurement spectra were recorded using a 7.8 Hz equivalent noise bandwidth and their peak gain frequency was estimated using a quadratic fitting. The repeatability for the HCF and the SMF is experimentally estimated to be 0.3 $^\circ$C and 0.4 $^\circ$C, respectively (see Supplementary Section S3.7 for additional details).
	
	Figure \ref{fig_Distributed}(b) shows the longitudinal distribution of the Brillouin frequency shift for various preset temperatures in the HCF. The slightly different positioning of the step transitions is due to the uncertainty in the central frequency determination when two Brillouin gain spectra overlap in presence of noise. The average Brillouin frequency shift in the HCF along the thermo-electric element as a function of the preset temperature is shown in Fig. \ref{fig_Distributed}(c). It should be pointed out that the response is not perfectly linear but shows an average slope of 1.2 MHz/$^{\circ}$C, which turns out to be conveniently slightly larger than for silica. The slope is higher for lower temperatures (in Fig. \ref{fig_Distributed}(c)), as a result of the closer vicinity to the liquefaction temperature ($\sim 8~ ^{\circ}$C at 41 bar), in agreement with a previous work \cite{estrada1998speed}.
	
	Figures \ref{fig_Distributed}(d) and (e) compare the response of each fibre at a preset temperature of 35$^\circ$C and under different applied strains: 0 $\rm {\upmu\upepsilon}$, 2000 $\rm {\upmu\upepsilon}$ and 4000 $\rm {\upmu\upepsilon}$. As expected, a strong strain dependence is observed for the SMF, but no change is visible for the HCF, validating the absence of observable cross-sensitivity, which was subsequently confirmed up to 1\% elongation. This experimental result consolidates the numerical simulations predicting this strain insensitivity (Supplementary Section S7). \\

	\noindent
	{\large{\textbf{Discussion}}}
	
	\noindent
	The large light-sound interaction in a gas-filled HCF reported here leads to a measured gain nearly six orders of magnitude larger than in previous works using free-space optics. In a pure fibre perspective, the increased compressibility of gases and their lower acoustic attenuation with respect to solid materials result in a measured Brillouin gain in our 41-bar \cod\ gas-filled HCF that turns 6 times larger than in standard SMF. 
	
	The issue of optical amplification in HCFs has already given rise to sustained efforts. Interesting results have been reported, mostly using molecular/atomic transitions in a low-pressure gas \cite{hassan_cavity-based_2016} (as in classical gas lasers) or Raman gain in hydrogen \cite{benabid_stimulated_2002}. The obtained gains remain modest in both cases when compared to solid-core solutions, with the specific penalties of amplification at fixed wavelengths for molecular/atomic transitions and the stringent issue of hydrogen permeation through the glass walls for Raman amplification. In contrast, our approach offers an efficient alternative to amplify signals since it shows a gain 20 times larger than the highest achievable Raman gain at 1.55 $\rm{\upmu m}$. Moreover, this amplification scheme operates at any wavelength from the ultraviolet to the mid-infrared region, limited only by the transmission windows of the HCFs.
	
	Gas Brillouin lasing has not yet been reported due to the extremely low scattering efficiency in free-space implementations. Using gas-filled HCF, we have demonstrated the first continuous-wave gas Brillouin laser with only 33 mW of threshold power, despite the high cavity roundtrip loss. The lasing threshold can be further decreased by dynamically matching the cavity resonance to the pump frequency in a doubly resonant configuration (both pump and Stokes are resonant). In addition, by changing the gas pressure, we evidenced that we can not only scale the gain but also modify the acoustic lifetime, which turns out to be a very important feature for building gas Brillouin photon or phonon lasers \cite{otterstrom_silicon_2018}. As clarified below, the nature of the gas is not very crucial, so that it can be perfectly possible to realise a gas Brillouin laser using compressed air as amplifying medium.
	
	In a different perspective, we also demonstrated a high-performance distributed temperature sensor showing spatial and temperature resolutions of 1.28 cm and 0.3 $^{\circ}$C, respectively. Note that, since we used a correlation-domain technique, the 10 times narrower linewidth (i.e. longer acoustic lifetime) compared to silica core SMF has no impact on the spatial resolution, but much improves the temperature resolution \cite{soto_modeling_2013}. Our distributed temperature sensor is robustly immune to high energy radiations (e.g. in space-borne situations or inside a nuclear reactor) where conventional solid-glass fibres are rapidly subject to photodarkening \cite{bykov_flying_2015}. The sensing range may be potentially extended to several tens of kilometres by using low-loss HCFs \cite{bradley_antiresonant_2019} filled with gases free of absorption in the C-band (such as nitrogen or noble gases).

\begin{figure}[tbp]
		\centering{
			\includegraphics[width = 1 \linewidth]{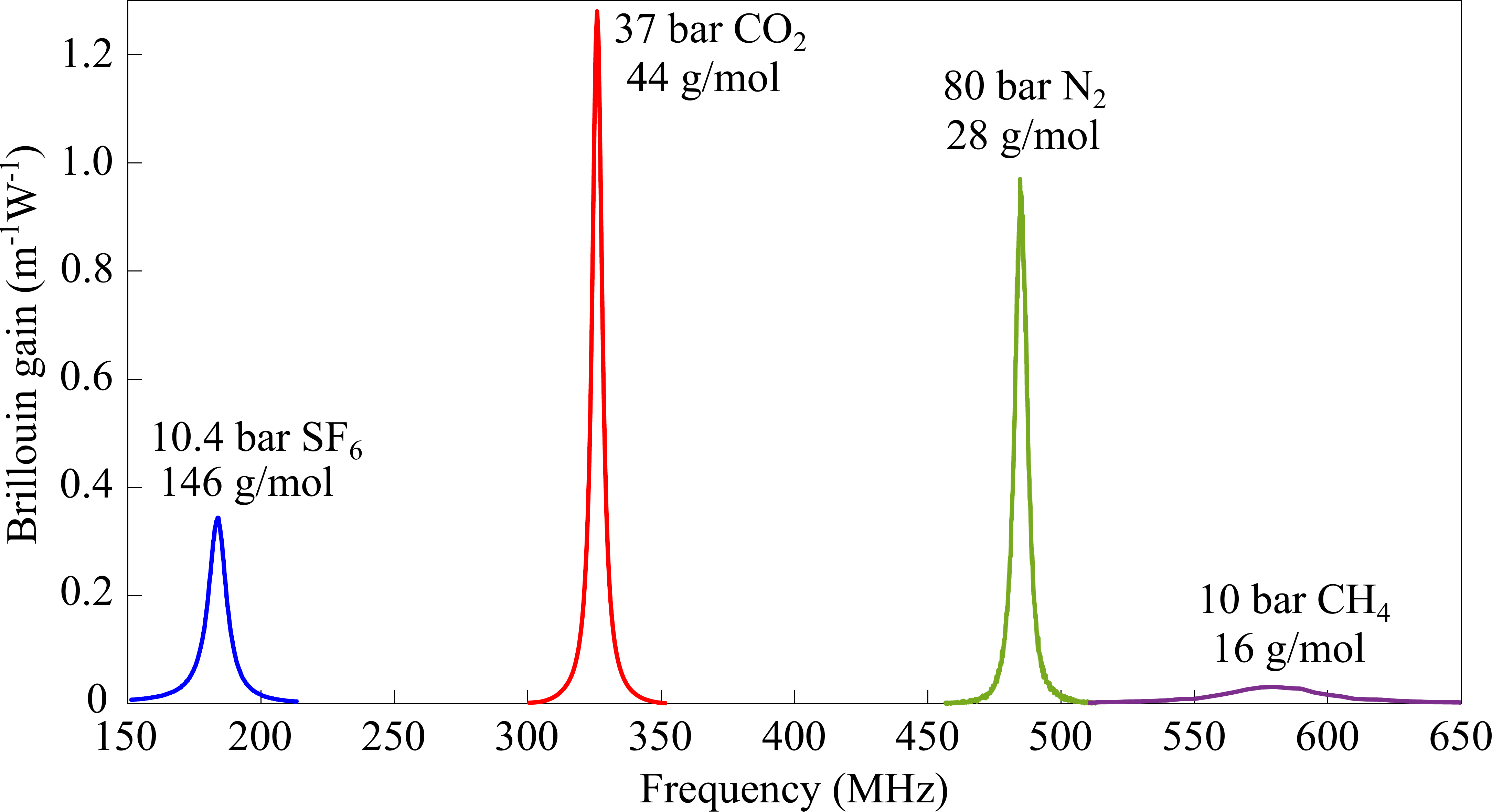}
		}
		\caption{\textbf{Experimental Brillouin gain in the HCF filled with different types of gas.} Measured Brillouin gain spectra for 37 bar CO$_2$, 80 bar N$_2$, 10.4 bar SF$_6$ and 10 bar CH$_4$. The molar mass of these four gases is also indicated in this figure. 
			\label{fig_DifferentGases}}
	\end{figure}

	The natural question arises as to how the nature of the gas influences the Brillouin gain. From our theoretical analysis it comes up straightforwardly that the gain depends quadratically on the gas density, so that heavier molecules are expected to present a larger amplification potential. We observed SBS in a HCF filled with different types of gas, namely carbon dioxide (\cod), sulfur hexafluoride ($\s{SF_6}$), nitrogen ($\s{N_2}$) and methane ($\s{CH_4}$). The Brillouin gain spectra for these four gases under specific pressures are plotted in Fig. \ref{fig_DifferentGases}. The acoustic velocity, and hence the Brillouin frequency shift, usually scales inversely to the square root of the gas molecular mass, as demonstrated in Fig. \ref{fig_DifferentGases}. \cod\ has been selected in this work for four main reasons: (1) its absorption at a wavelength of 1.55 $\rm {\upmu m}$ remains limited below 41 bar, (2) it has a relatively large density, leading to a Brillouin gain 6 times and 3 times larger than $\s{N_2}$ and $\s{CH_4}$ respectively at the same pressure, (3) compared to SF$_6$, \cod\ shows a higher liquefaction pressure at room temperature, which results in a higher achievable gain (even though the Brillouin gain of \cod\ is 3 times lower than that of $\s{SF_6}$ at the same pressure (e.g. 10 bar)) and (4) it is widely available, does not permeate through glass and can be handled with no potential hazards.

	In our experiments, the maximum pressure used for \cod\ is 41 bar, which is not a physical limitation of the HCF, since such fibres can easily sustain a pressure in the kilobar range by virtue of their small core diameter and thick glass sheath \cite{russell_hollow-core_2014}. This maximum pressure turns out to lie below the onset of substantial light absorption at 1.55 $\upmu$m. As observed at pressures above 41 bar, the substantial light absorption due to pressure broadening impairs the HCF transmission and hence decreases the Brillouin signal. As a result, it has to be mentioned that the use of complex heavy molecules is eventually of limited relevance, since such molecules normally present broad and numerous absorption bands and frequently liquefy at moderate pressure (e.g. 61 bar for \cod\ and 22 bar for $\s{SF_6}$) at room temperature. Using simpler molecules such as nitrogen, oxygen or noble gases opens the possibility to raise the pressure without risk of liquefaction, so that their smaller intrinsic gain can be eventually much overcompensated by a higher pressure. Such gases are normally also totally free of spectral absorption lines in the regions of interest. For instance a theoretical gain up to 30 m$^{-1}$W$^{-1}$ is anticipated in a xenon-filled HCF for a pressure above 130 bar. This gain is more than 100 times larger than in a solid-core silica fibre.
	
	Our platform is also suitable for the investigation of light-sound interactions in gases close to their critical point or in the supercritical region, as well as for the study of the bulk viscosity at high frequency, which is, so far, poorly documented. It should be pointed out that the gas consumption in volume for a 50 m long HCF having its core and cladding filled with 1 kilobar gas is equivalent to only about 200 ml at atmospheric pressure, thanks to the microscopic size of the structure. In a more practical approach, the gas-filled HCF can be hermetically sealed by splicing both ends to standard SMFs, thereby making a perfectly airtight compact all-fibre gas cell \cite{benabid_compact_2005} which can be flexibly and safely handled. 
	
	Stimulated Brillouin scattering cannot be reduced to a mere amplification process, since it has demonstrated its potentialities to realise advanced functions \cite{santagiustina_all-optical_2013}. This novel gas-based Brillouin platform can be the foundation of many potential applications, some being partially illustrated here: amplifiers, highly coherent Brillouin gas lasers, slow \& fast light, microwave filters, tuneable delay lines, light storage, all-optical calculus and of course sensing. The same functions can therefore be implemented in hollow-core fibres, offering all the inherent assets of fibre-based optics, with the key advantage to realise the same response with a product \textit{pump power} $\times$ \textit{fibre length} potentially 100$\times$ smaller. It must be mentioned that the reduced acoustic loss with respect to silica results in a narrower gain resonance: this may be seen at first glance only as a drawback since it reduces the capacity for broadband amplification. However, it turns out to be a clear asset for the majority of applications benefiting from a long-lasting vibration: optical storage, optical signal processing, precisely selective spectral filtering, sensing, etc...
	
	On a broader perspective, the concept introduced in this paper can also be applied to other waveguiding structures. More specifically, although suspended silicon or silica waveguides can be designed to exhibit a light-sound interaction \cite{eggleton_brillouin_2019}, the interaction between the evanescent field of their guided light and the gas has not yet been exploited and could lead to gains of practical interest. For instance, a small dimension slot waveguide, inducing an intense evanescent field \cite{almeida_guiding_2004} can be designed to offer a large light-sound interaction in gas. This shows that, if the immediate and massive benefit of Brillouin amplification in gas is undoubtedly for hollow core waveguides, its potentialities can certainly extend to other configurations.
	
	\vspace{0.5cm}

	\noindent
	\textbf{Methods}
	
	\noindent
	\textbf{Fabrication of the HCF gas cell}
	
	\noindent
	{\small{Thanks to the similar core and cladding diameters of our HCF and a standard SMF, the HCF gas cell can be formed by placing a segment of HCF between two SMF patchcords according to the following procedure: (1) two ceramic fibre ferrules having an inner diameter of 125 $\rm{\upmu m}$ were inserted into a ceramic sleeve, keeping a 30 $\rm{\upmu m}$ gap between the two ferrule tips. The ceramic sleeve's side slot enables visual monitoring of the butt coupling at fabrication stage and gas inlet/outlet into/from the HCF under operation. (2) An angled-cleaved SMF and a cleaved end of the HCF were inserted into the fibre ferrules and the coupling of the HCF/SMF was monitored by a microscope through the side slot. HCF and SMF ends were brought closer to each other until they are separated by a few $\rm{\upmu m}$ gap. The other end of the HCF was coupled according to the same procedure. The total end-to-end loss for the assembling SMF/HCF/SMF is 9 dB. (3) We inserted each joint into a metallic T-tube and sealed its two facing sides using epoxy glue. Gas can be vacuum-pumped out or pressurised into the HCF through the third port of the T-tubes.}} \\
	
	\noindent
	\textbf{Simulations}
	
	\noindent
	{\small{Simulations of the optical and acoustic modes are performed using COMSOL Multiphysics 2D "Electromagnetic Waves" and 2D "Pressure Acoustics" modules, respectively. The silica refractive index $n_{\rm SiO_2}$=1.444 and the gas refractive index $n_{\rm gas}$=1.01804 (for 41 bar \cod) are entered into the calculation of the optical modes. The effective refractive index of the fundamental optical mode $n_{\rm eff}$ is calculated to be 1.0123 at a wavelength of 1.55 $\upmu \rm m$. The fundamental mode profile (optical intensity) is plotted in Fig. \ref{fig_Introduction}(b). At 41 bar \cod, the gas density $\rho_{\rm gas}=72.77$~$\rm kg/\rm m^3$ (ideal gas approximation) and the speed of sound $v_{\rm a}=243.6~\rm m/\rm s$ (as deduced from the Brillouin frequency shift of our measurement) are used to calculate the acoustic mode in the fibre core, considering a sound hard boundary on the hollow tube wall. The acoustic mode profile (density) of the calculated first excited radial mode is shown in Fig. \ref{fig_Introduction}(c) with an out-of-plane wave-vector of $8.228\times10^6$ rad/m at a resonant frequency of 320 MHz, which corresponds to the measured Brillouin frequency shift at 41 bar \cod.}} \\

	\noindent
	{\small \textbf{Acknowledgments:} We acknowledge support from the Swiss National Foundation under grant agreements No. 178895 and 159897. We thank Dr. Meng Pang from Shanghai Institute of Optics and Fine Mechanics for fruitful and valuable discussions, Fu Yun, Suneetha Sebastian and Benjamin Pickford for the revision of this manuscript.}
	
	\noindent
	{\small \textbf{Author contributions:} L.T. initiated the idea of exploiting SBS in gases through hollow-core fibres. F.Y. conceived the idea of giant Brillouin amplification by using pressurised gas in HCFs. L.T. conceived the strain-insensitive sensing idea. F.Y. and F.G. fabricated the HCF gas cell, designed the measurement set-ups and performed the experiments. F.Y. and F.G. simulated the acoustic and optical modes and theoretically analysed the gain coefficient. F.G. explained the acoustic attenuation in relation to the gas pressure, simulated the impact of strain on the gas-filled HCF. F.Y. and F.G. wrote the manuscript with inputs from L.T.. L.T. supervised this work.}
	
	\noindent
	{\small \textbf{Data availability:} The code and data used to produce the plots within this work will be released on the repository \texttt{Zenodo} upon publication of this preprint.}

	\bibliographystyle{spiebib}   
	\bibliography{main_ref}
	
\end{document}


\maketitle

\tableofcontents

\section{Dual intensity modulation}
\label{sec:dualintensitymodulation}
\subsection{Introduction}
\begin{figure}[htbp]
	\centering
	\includegraphics[width=0.7\linewidth]{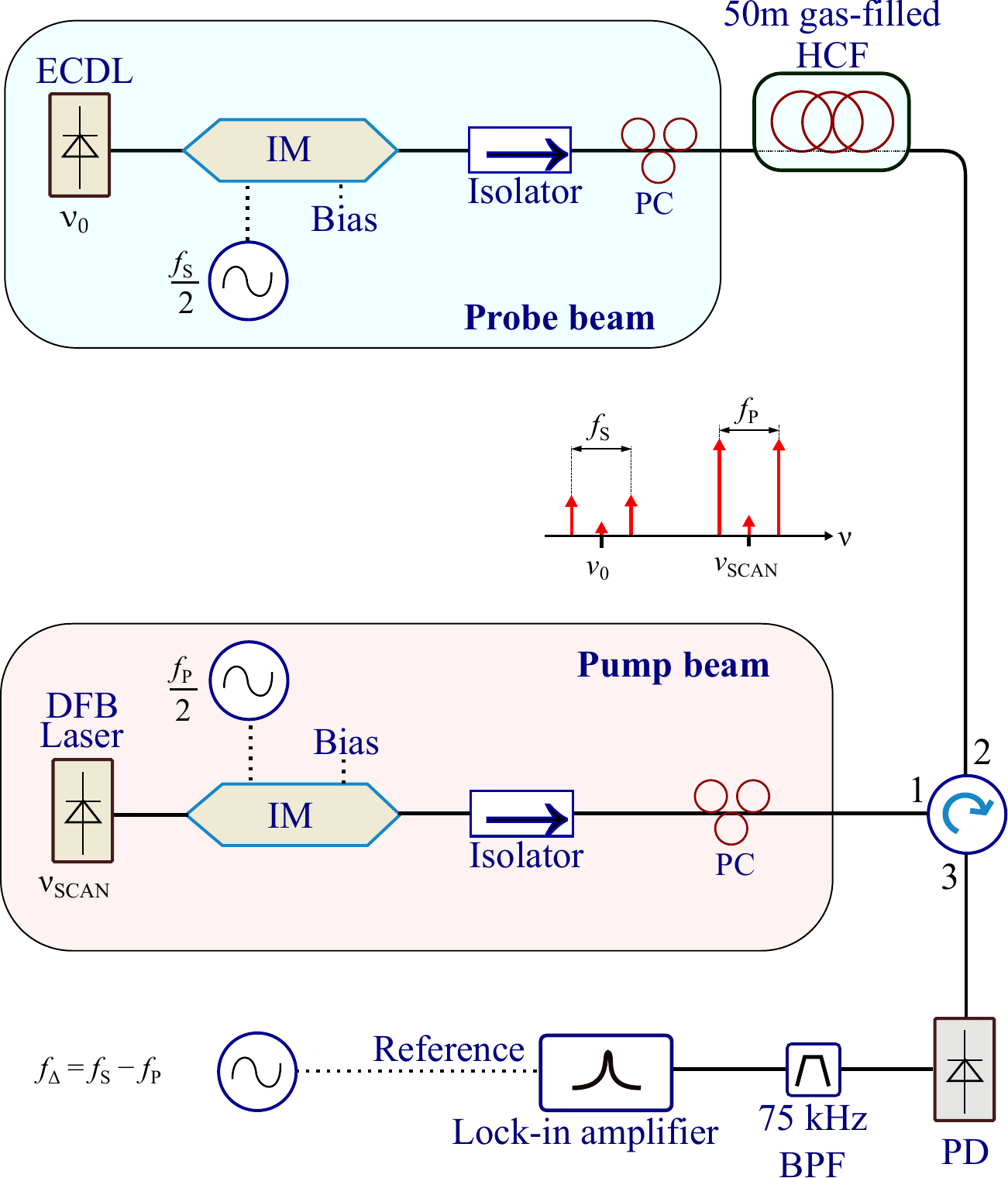}
	\caption{\textbf{Dual intensity modulation (DIM) experimental set-up.} Experimental set-up for Brillouin gain spectrum measurement using an intensity modulation for both pump and probe beams. Detection is performed at the frequency difference. Note that the three radio-frequency sources are synchronised to the same frequency standard. ECDL, external-cavity diode laser; DFB laser, distributed feedback laser; IM, intensity modulator; PC, polarisation controller; PD, photodetector; BPF, band-pass filter.}
	\label{fig:double_mod_schematic}
\end{figure}

When measuring the Brillouin gain in low-pressure gas, the peak gain is relatively low (e.g. in 1 bar \cod, it is about $10^{-3}~\s{m^{-1}W^{-1}}$). In this situation, the pump reflection at the SMF- HCF coupling interface, directly entering into the detector, is the source of fluctuations screening the gain to be measured. This reflection issue is resolved by introducing a dual intensity modulation at frequency $f_S/2$ and $f_P/2$ on the probe and pump beam, respectively. The modulators are Mach-Zehnder modulators and their bias is set to suppress the carrier. As a consequence, the probe and pump light intensities are modulated at a frequency $f_S$ and $f_P$, respectively. Since \sbs\ is a non-linear process involving the product of pump and probe powers, sum and difference of frequencies are generated. After Brillouin interaction with the pump inside the gas-filled hollow-core fibre (HCF), the probe intensity is detected and band-pass filtered at a frequency $f_\Delta = f_S - f_P$ using a lock-in amplifier. Hence, the pump beam directly reaching the detector is filtered out in the radio frequency (RF) domain, since the detection is made at a frequency very distant from $f_P$ \footnote{However, the reflection still needs to remain limited as to not damage the photodetector in the case of a high pump power. Furthermore, the photodetector may show a slight nonlinearity. In that case, the presence of both pump and probe beams at the detector could lead to sum-frequency difference generation within the detector itself, generating a background noise that may cover the desired signal. In our case, angled-cleaved SMFs are used to sufficiently reduce the reflection.}. The experimental set-up is illustrated in Fig. \ref{fig:double_mod_schematic}. Note that the frequency difference, $f_\Delta$, should remain much smaller than the Brillouin linewidth to secure steady-state acoustic waves. A similar technique is used to filter out stray light in Brillouin microscopy \cite{grubbs1994high}.

\subsection{System response}
\label{sec:dualmodulation_derivation}
In this derivation, it is assumed that both intensity modulation frequencies $f_S$ and $f_P$ are much smaller than the Brillouin frequency shift $\nu_B$. Moreover, the frequency difference $f_P - f_S$ is assumed to be much smaller than the Brillouin linewidth $\Gamma_B/2\pi$. Then, the acoustic wave governing the interaction at frequency $f_\Delta$ can be approximated to be at steady-state. Furthermore, absence of pump depletion is assumed. In these conditions, the probe electric field along the HCF (z-axis), $E_s(z)$, in presence of \sbs, is governed by the following differential equation \cite{boyd_book_sectionBrillouin}:

\begin{equation}
\frac{\partial E_s}{\partial z}+ \frac{n_g}{c}\frac{\partial E_s}{\partial t} =\frac{1}{2}g_B \mathcal{L}_A P_P E_s -\frac{\alpha}{2}E_s,
\label{diff_equa_sm1}
\end{equation}
where $g_B$ is the peak Brillouin gain, as defined in Eq. (1) in the main manuscript, $c$ is the speed of light, $n_g$ is the group refractive index of the fibre, $\alpha$ is the fibre loss in unit 1/m, $P_P(z) = P_{P0}\ex{-\alpha (L-z)}$ is the pump power along the HCF, $P_{P0}$ being the input pump power at $z = L$, where $L$ is the fibre length, and 
\begin{equation}
\mathcal{L}_A(\Omega) = \frac{\i\Omega\Gamma_B}{\Omega_B^2-\Omega^2+\i\Omega\Gamma_B}
\end{equation}
is the probe (amplification) field lineshape, for which the \ppfd\ $\Omega/2\pi = \nu_{\s{SCAN}}-\nu_0$, the \bfs\, $\nu_B = \Omega_B/2\pi$, and the Brillouin linewidth (full width at half maximum), $\Delta\nu =\Gamma_B/2\pi$, are defined.

The equation is converted in units of power by using $P_S = A_\s{eff}\|E_s\|^2/2\eta$, where $A_\s{eff}$ is the fibre effective area and $\eta$ is the gas impedance:

\begin{equation}
\begin{split}
\frac{1}{P_S}\Biggl(\frac{\partial P_S}{\partial z}+\frac{n_g}{c}\frac{\partial P_S}{\partial t} \Biggr)=g_B \mathcal{L}_I P_P- \alpha,
\end{split}
\label{diff_equa_sm2}
\end{equation}
where $\mathcal{L}_I(\Omega) = \norm{\mathcal{L}_A(\Omega)}^2$ is the intensity lineshape. In order to solve Eq. (\ref{diff_equa_sm2}), we change the coordinate frame from ($z$, $t$) to ($z'$, $t'$) by performing the following change of variable:
\begin{equation}
\begin{split}
&z\left(z', t'\right) = z',\\
&t(z', t') = t' + \frac{n_g}{c}z'.
\end{split}
\end{equation}
Using these expressions for the original coordinates as a function of the new coordinates, we apply the chain rule and express:
\begin{equation}
\begin{split}
&\frac{\partial P_S(z, t)}{\partial z'} = \frac{\partial P_S(z, t)}{\partial z}\frac{\partial z}{\partial z'} + \frac{\partial P_S(z, t)}{\partial t}\frac{\partial t}{\partial z'} = \frac{\partial P_S(z, t)}{\partial z} + \frac{n_g}{c}\frac{\partial P_S(z, t)}{\partial t},\\
&\frac{\partial P_S(z, t)}{\partial t'} = \frac{\partial P_S(z, t)}{\partial z}\frac{\partial z}{\partial t'} + \frac{\partial P_S(z, t)}{\partial t}\frac{\partial t}{\partial t'} = \frac{\partial P_S(z, t)}{\partial t}.
\end{split}
\end{equation}

These expressions allow rewriting the partial differential equation, Eq. (\ref{diff_equa_sm2}), as:
\begin{equation}
\frac{1}{P_S(z, t)}\frac{\partial P_S(z, t)}{\partial z'} =g_B \mathcal{L}_I P_P(z', t')- \alpha,
\label{diff_equa_bossted}
\end{equation}
where the pump power $P_P$ is now expressed as a function of the new coordinates. Since the time derivative disappears, it is now easier to solve the equation. An expression for the the pump power generated by the Mach-Zehnder modulator, $P_P(z,t)$, is derived in section \ref{appendix_intensity_modulation}. Using this derivation and the fact that the modulator bias is in carrier-extinct configuration, $P_P$ can be expressed as:
\begin{equation}
P_P(z,t) = \tilde{P}_{P0}\ex{-\alpha (L-z)}\Biggl(1-\sum_{n}J_n(2\varsigma_P)\cos{\bigl(n(K_P z -\Omega_P t)\bigr)}\Biggr),
\end{equation}
where $P_{P0} = \tilde{P}_{P0}\left(1-J_0(2\varsigma_P)\right)$ is the time-averaged input pump power at the end of the HCF ($z = L$), $\Omega_P = \pi f_P$ and $K_P = -\Omega_Pn_g/c$. We now transform this expression from ($z$, $t$) to the new coordinate frame, ($z'$, $t'$). Given that the pump counter-propagates,  $K_P<0$, and the expression becomes:
\begin{equation}
P_P(z',t') = \tilde{P}_{P0}\ex{-\alpha (L-z')}\Biggl(1-\sum_{n}J_n(2\varsigma_P)\cos{\bigl(n(2K_P z' -\Omega_P t')\bigr)}\Biggr),
\label{singlepump_pumppowerexpression}
\end{equation}
which, to simplify subsequent integration, can also be expressed as:

\begin{equation}
P_P(z',t') = \tilde{P}_{P0}\left(\ex{-\alpha (L-z')}-\ex{-\alpha L}\Re\left\lbrace\sum_{n}J_n(2\varsigma_P)\ex{(\alpha +2\i nK_P) z' -\i n\Omega_P t'}\right\rbrace\right).
\label{singlepump_pumppowerexpression_simplifyintegration}
\end{equation}
Replacing pump power expression (\ref{singlepump_pumppowerexpression_simplifyintegration}) into (\ref{diff_equa_bossted}) and integrating both sides along $z'$ yields:
\begin{equation}
\begin{split}
&\ln{\left\lbrace P_S\left(z(z', t'), t(z', t')\right)\right\rbrace}\bigg\vert^{z' = L}_{z' = 0} = \int_{0}^{L}\left(g_B \mathcal{L}_I P_P(z',t')- \alpha\right) \dd z'\\
&= \left(g_B \mathcal{L}_I \tilde{P}_{P0}\left(\frac{1}{\alpha}\ex{-\alpha(L-z')}-\ex{-\alpha L}\Re\left\lbrace\sum_{n}J_n(2\varsigma_P)\frac{\ex{(\alpha +2\i nK_P) z' -\i n\Omega_P t'}}{\alpha +2\i nK_P}\right\rbrace\right) -\alpha z'\right)\bigg\vert^{z' = L}_{z' = 0}.
\end{split}
\end{equation}
Hence we obtain the general solution: 
\begin{equation}
P_S(L, t) = P_{S}(0, t-n_gL/c)\ex{-\alpha L}\exp{\left(\tilde{P}_{P0}g_B \mathcal{L}_I L_\s{eff}\left(1-\Re\left\lbrace \sum_{n}J_n(2\varsigma_P)\zeta_n\ex{-\i n\Omega_P t}\right\rbrace \right)\right)},
\label{single_mod_Ps_equa} 
\end{equation}
where
\begin{equation}
L_{\s{eff}}= \frac{1-\ex{-\alpha L}}{\alpha}
\label{eq:effectivelength}
\end{equation}
is the fibre's effective length and

\begin{equation}
\zeta_n = \frac{1}{L_\s{eff}}\frac{\alpha -2\i nK_P}{\alpha^2+(2nK_P)^2}\bigl(\ex{\i n K_P L}-\ex{-\i n K_P L -\alpha L}\bigr)
\end{equation}
is a modulation-dependent unitless parameter. The input probe power can be expressed as:

\begin{equation}
P_{S}(0, t-n_gL/c) = \tilde{P}_\s{S,0}\biggl(1-\sum_{m}J_m(2\varsigma_S)\cos\bigl(m\Omega_S t + m\Phi_s\bigr)\biggr),
\end{equation}
where $\Omega_S = \pi f_S$ is the modulation angular frequency, $P_\s{S,0} = \tilde{P}_\s{S,0}\left(1-J_0(2\varsigma_S)\right)$ is the time-averaged input probe power at z = 0, $\Phi_s$ contains both a phase shift between pump and probe modulations as well as the phase shift due to the probe propagation along the fibre. $\varsigma_S$ is the probe modulation depth. Section \ref{appendix_intensity_modulation} gives a detailed derivation of intensity-modulated signals using a Mach-Zehnder modulator, including the definition of the modulation depth, $\varsigma$. Since this technique was developed for the measurement of small Brillouin gains, we can now apply the small-gain approximation. After applying this approximation, the obtained expression consists of:
\begin{itemize}
	\item A DC term.
	\item A term oscillating at multiples of the probe frequency, $f_S$.
	\item A term oscillating at multiples of the pump frequency, $f_P$.
	\item A term oscillating at multiples of the difference between the pump and probe frequencies, $f_S-f_P$.
\end{itemize}
As the detection is performed at the difference of the frequencies, only the last term is considered. The power in the small-gain approximation can then be written as:
\begin{equation}
P_{S,(\Omega_P\Omega_S)}(L, t) \cong -\tilde{P}_\s{S,det}\tilde{P}_{P0}g_B \mathcal{L}_I L_{\s{eff}}\Re\left\lbrace \sum_{n}\sum_{m}J_n(2\varsigma_P)J_m(2\varsigma_S)\zeta_n\ex{-\i n\Omega_P t}\cos{\left(m\Omega_S t+m\Phi_S\right)}\right\rbrace,
\end{equation}
where $P_\s{S,det} = \tilde{P}_\s{S,det}\left(1-J_0(2\varsigma_S)\right)$ is the time-averaged probe power at detection. Since the lock-in amplifier only responds to the frequency difference $\Omega_\Delta = 2\left(\Omega_S-\Omega_P\right)$, only the terms $n = \pm 2,m=\pm 2$ are kept:
\begin{equation}
\begin{split}
P_{S,\Omega_\Delta}(L, t) \cong& -2\tilde{P}_{S,\s{det}}\tilde{P}_{P0}g_B \mathcal{L}_I L_\s{eff} J_2(2\varsigma_S)J_2(2\varsigma_P)\\
\cdot\biggl(&\cos{\left(\Omega_\Delta t\right)}\left(\Re{\left\lbrace\zeta_2\right\rbrace}\cos{(2\Phi_S)} +\Im{\left\lbrace\zeta_2\right\rbrace}\sin{(2\Phi_S)}\right)\\
+&\sin{\left(\Omega_\Delta t\right)}\left(\Re{\left\lbrace\zeta_2\right\rbrace}\cos{(2\Phi_S)} +\Im{\left\lbrace\zeta_2\right\rbrace}\sin{\left(2\Phi_S\right)}\right)\biggr).
\end{split}
\end{equation}
The photoreceiver delivers a voltage equal to $V_\s{det} = \rho_{\s{pd}} P_\s{det}$, where $\rho_{\s{pd}}$ is the power-to-voltage conversion factor. Then, in presence of a voltage signal of type $A\cos(\Omega t) + B\sin(\Omega t)$, the lock-in detection outputs the magnitude $\sqrt{A^2+B^2}$, which gives:

\begin{equation}
V_{s,\s{det}} \cong \frac{2J_2(2\varsigma_S)J_2(2\varsigma_P)}{\left(1-J_0(2\varsigma_S)\right)\left(1-J_0(2\varsigma_P)\right)}\rho_{\s{pd}}P_{S,\s{det}}P_{P0}g_B \mathcal{L}_I L_\s{eff}\norm{\zeta_2}.
\label{eq:double_mod_response_final}
\end{equation}
The dependence of the set-up's response on the length and modulation frequency is given by the parameter:
\begin{equation}
\norm{\zeta_n} = \frac{1}{L_\s{eff}}\sqrt{\frac{1+\ex{-2\alpha L}-2\ex{-\alpha L}\cos{\left(2nK_P L\right)}}{\alpha^2+(2nK_P)^2}}.
\label{eq:zeta_norm}
\end{equation}
The parameter $\zeta_n$ quantifies the interference effect occurring when the wavelength related to the modulation frequency $f_P$ becomes comparable or smaller than the effective length of the fibre. That is, when the following condition: $\Lambda < L_\s{eff}$, where $\Lambda = c/(n_gf_P)$, is met, the system's response significantly decreases. On the contrary, when $\Lambda \gg L_\s{eff}$, the parameter $\zeta_n\approx 1$. Note that the pump beam is also attenuated by the probe beam. As a result, a $\pi$-phase shifted signal, co-propagating with the pump beam and of same magnitude as the one described above, is also generated. If the reflection of the HCF is too large (> $-20$ dB), then the reflection of this signal will interfere significantly with the main signal and this can be no longer neglected. In our case, the angled-cleaved SMF and the HCF connector generate <$-36$ dB reflection. We can therefore neglect the contribution from the pump reflection. Finally, note that in the special case of a bias at the quadrature-point for the two modulators (as used for the distributed temperature experimental set-up, section \ref{sec:dts_sensing}), the result given by Eq. (\ref{eq:double_mod_response_final}) is modified to:
\begin{equation}
V_{s,\s{det,QP}} \cong 2J_1(2\varsigma_S)J_1(2\varsigma_P)\rho_{\s{pd}}P_{S,\s{det}}P_{P0}g_B \mathcal{L}_I L_\s{eff}\norm{\zeta_1}.
\label{eq:double_mod_response_final_QP}
\end{equation}

\subsection{Experimental verification}
Here, we experimentally verify the set-up response provided by Eq. (\ref{eq:double_mod_response_final}) and compare it with a single modulation set-up.
\begin{figure}[htbp]
	\centering
	\includegraphics[width=\linewidth]{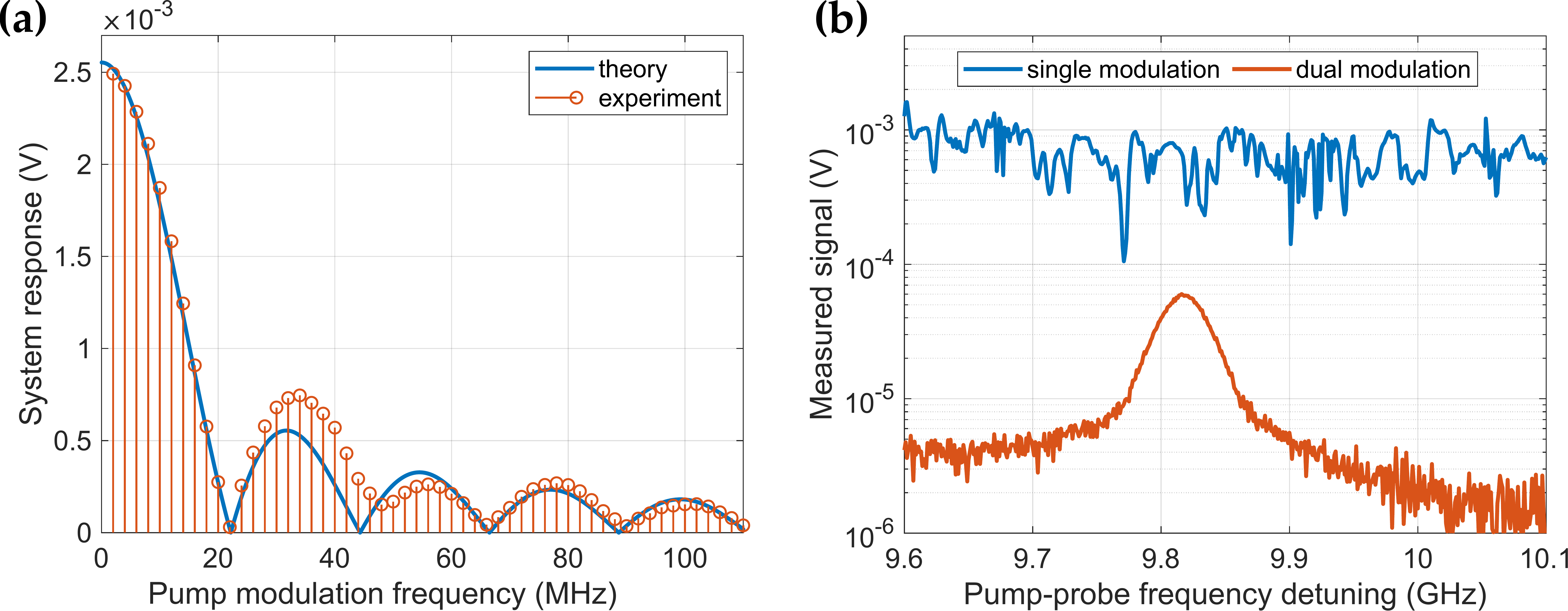}
	\caption{\textbf{Experimental verification of the dual intensity modulation method.} \textbf{(a)} Comparison of the measured system response with the calculated response given by Eq. (\ref{eq:double_mod_response_final}), showing good qualitative agreement. \textbf{(b)} Comparison of the performances with the single modulation set-up in presence of a $-$48 dB reflection at the fibre coupling interfaces. In the case of the single intensity modulation method, the Brillouin gain signal is fully screened by the background fluctuations due to the pump reflection. By contrast, the signal from the dual intensity modulation method is not perturbed by this reflection and is able to accurately measure the gain.}
	\label{fig:dual_mod_expverification}
\end{figure}
\subsubsection{System response}

A 5-m-long standard \ssmf\ single-mode fibre was connected to the experimental set-up shown in Fig. \ref{fig:double_mod_schematic} (replacing the gas-filled HCF with the 5-m-long SMF). Both pump and probe modulation frequencies were gradually increased from 2 MHz to 110 MHz by 2 MHz steps, such that their difference, $f_\Delta$, remained constant and equal to 75 kHz for the entire sequence. For each frequency step, the Brillouin gain was acquired and its peak value recorded. Figure \ref{fig:dual_mod_expverification}(a) plots the results as a function of the pump modulation frequency, together with the theoretical curve obtained from Eq. (\ref{eq:double_mod_response_final}), assuming a Brillouin gain of $g_B$ = 0.25 \mw. We can observe that the measurements match well with the theoretical curve. Small deviations are observed and likely come from the presence of the connecting patchcords showing a slightly different \bfs .

\subsubsection{Robustness to reflections}
In order to verify the robustness of the dual intensity method against reflections, a 52-cm-long small-solid core fibre (ultra-high NA fibre with 1.8 $\upmu$m core diameter), with its both ends spliced to a single-mode fibre patchcords was used (i.e. replacing the gas-filled HCF with the small-core fibre in Fig. \ref{fig:double_mod_schematic}). This sample exhibits a reflection of $-$48 dB at the coupling interfaces caused by the effective index mismatch between the two fibre types. The gain of the second acoustic mode (high-order acoustic mode) of the small core fibre ($g_B \approx$ 0.062 \mw) was measured using the single modulation set-up and the dual modulation set-up with the same parameters (same pump power, probe power and modulation depth). The time-averaged pump power, just before entering the sample, was 14 dBm and the detected probe power (time-averaged for the dual modulation case) was $-$8 dBm. The results are shown in Fig. \ref{fig:dual_mod_expverification}(b). We can see that in the case of the single modulation, the pump reflection ($\approx~-$34 dBm) reaching the detector is sufficient to generate background fluctuations fully screening the Brillouin gain signal. On the other hand, in the case of the dual modulation set-up, the reflection is filtered out and the Brillouin gain can be measured with a good SNR.

\subsection{Experimental Brillouin gain calculation}
Here, we use Eq. (\ref{eq:double_mod_response_final}) to compute the Brillouin gain. The parameters are listed in Table \ref{table_characterisation_Brillouin_gain}.

\begin{table}[h]
\centering
\begin{tabular}{lll} 
Parameter name & Parameter description & Value\\
 \hline
 $\varsigma_P$ & Pump modulation depth & 1.15 \\ 
 $\varsigma_S$ & Probe modulation depth & 1.15 \\ 
 $P_{P0}$ & Pump power at the input of the HCF (inside, one sideband only)& 7.8 dBm\\
 $P_{S,\s{det}}$ & Probe power at the photodetector & -7.5 dBm\\
 $f_P/2$ & Pump modulation frequency & 714.623 kHz\\
 $f_S/2$ & Probe modulation frequency & 752.123 kHz\\
 $f_\Delta$ & Detection frequency & 75 kHz\\
 $\alpha$ & HCF optical attenuation (including the 0.5 dB \cod\ absorption) & 5.99 km$^{-1}$\\
 $n_g$ (41 bar)&HCF group refractive index&1.01804\\
 $\rho_\s{pd}$ & Photodetector power-to-voltage conversion factor & 3.75 V/mW\\
 $\alpha_F$ & Voltage attenuation factor due to the band-pass filter& 0.827\\
 $V_{s,\s{det}}$ (41 bar)&Lock-in amplifier voltage corresponding to the peak Brillouin gain& 60.8 mW\\
 \hline
\end{tabular}
\caption{\textbf{Parameters used for computation of the Brillouin gain from the dual intensity modulation system response.} Note that the values of $n_g$ and $V_{s,\s{det}}$ are given at 41 bar \cod-filled HCF as an example, and the pump and probe power are the time-averaged values.}
\label{table_characterisation_Brillouin_gain}
\end{table}

Using these values along with Eqs (\ref{eq:effectivelength}) and (\ref{eq:zeta_norm}), we can compute $L_\s{eff} = 43.21$ m, $\norm{\zeta_2} = 0.657$. We now add $\alpha_F$ to Eq. (\ref{eq:double_mod_response_final}) and rewrite it to obtain the expression of the Brillouin gain: 
\begin{equation}
g_B = \frac{\left(1-J_0(2\varsigma_S)\right)\left(1-J_0(2\varsigma_P)\right)\cdot V_{s,\s{det}}}{ 2J_2(2\varsigma_S)J_2(2\varsigma_P)\rho_{\s{pd}}P_{S,\s{det}}P_{P0} L_\s{eff}\alpha_F\norm{\zeta_2}}.
\end{equation}
By evaluating this equation using values for all parameters, we obtain $g_B = 1.68$ m$^{-1}$W$^{-1}$ for 41 bar \cod, which is in good agreement with our Brillouin amplification and lasing measurements.

\section{Detailed experimental set-up for signal amplification}

\begin{figure*}[htbp]
  \centering{
  \includegraphics[width = 0.8 \linewidth]{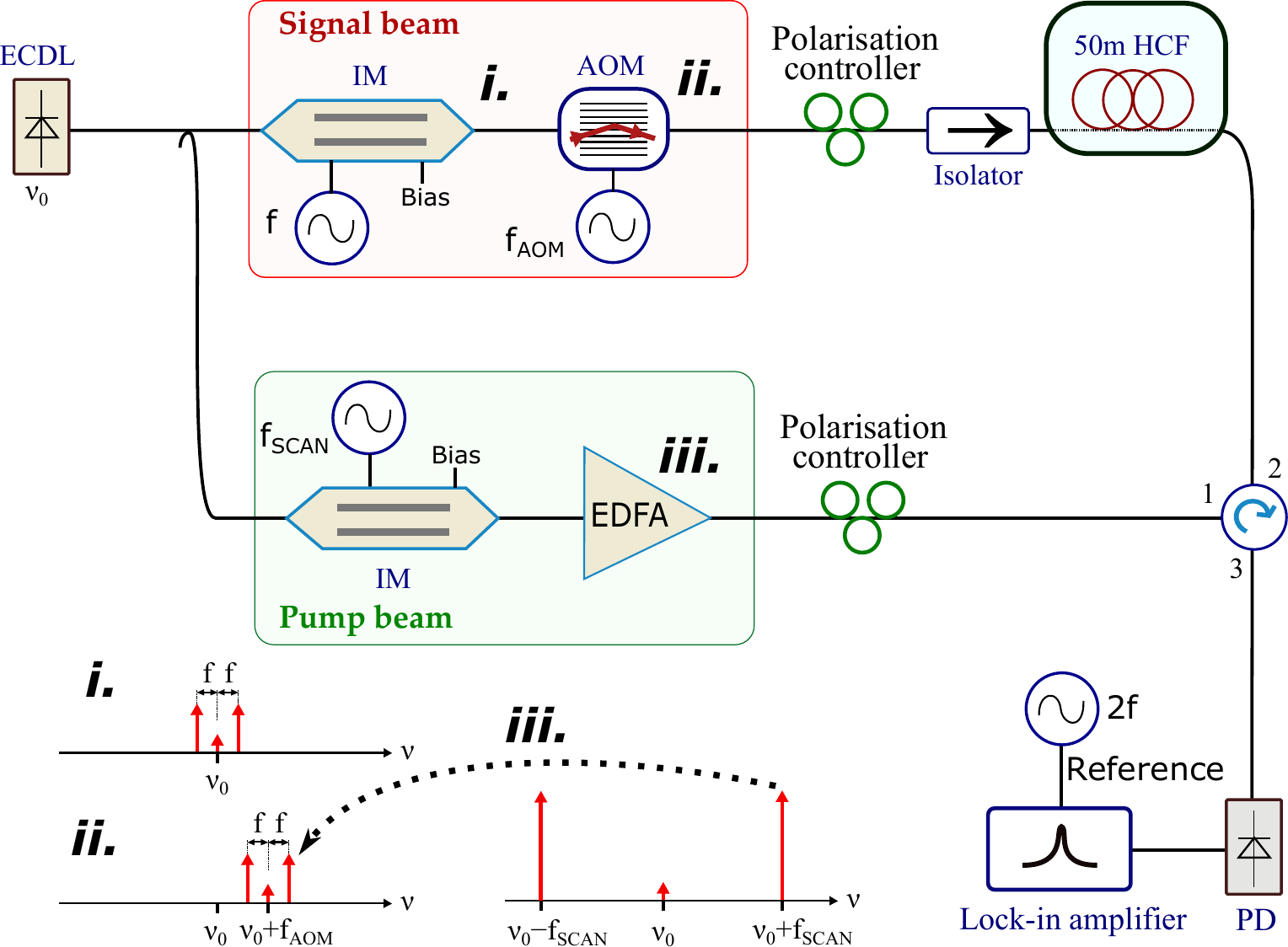}
  }
  \caption{\textbf{Detailed experimental set-up for the signal amplification measurement.} ECDL, external-cavity diode laser; IM, intensity modulator; AOM, acousto-optic modulator; EDFA, erbium-doped fibre amplifier; PD, photodetector. 
}
\label{fig:signal_amplification_setup}
\end{figure*}

\noindent
The detailed experimental set-up used for signal amplification measurements is shown in Fig. \ref{fig:signal_amplification_setup}. Pump and probe (signal) beams are both from the same external-cavity diode laser. The pump light was generated by a Mach-Zehnder electro-optic modulator (with carrier-suppressed bias set-up) at modulation frequency $f_{\rm SCAN}$ and then amplified by an erbium-doped fibre amplifier. The signal goes through an intensity modulator at frequency $f$=37.5 kHz (carrier-suppressed bias, hence the power is modulated at frequency $2f$ = 75 kHz) and then is frequency blue-shifted by an acousto-optic modulator ($f_{\rm AOM}=+110$ MHz), in order to break the pump sideband symmetry. Therefore, only one pump sideband is used for the Brillouin interaction (here, only the higher-frequency sideband of the pump beam interacts with the signal). The injected signal power (average power) is $-$34 dBm, which is more than 40 dB smaller than the pump power, hence satisfying the small-signal amplification condition (i.e. absence of pump depletion). 

By scanning the detuning frequency ($f_{\rm SCAN}-f_{\rm AOM}$) across the Brillouin frequency shift, we can measure the Brillouin amplification spectra as a function of the pump-signal detuning frequency using different pump powers. Here, the signal modulation frequency, $2f$, is much smaller than the linewidth of the Brillouin gain spectrum at 41-bar \cod\ (i.e. 3.65 MHz). This means that the two probe sidebands have the same Brillouin amplification coefficient. The output signal (inside the HCF before entering the output SMF) can be expressed as: $P_S(L) = P_{S}(0)\exp{\left(P_{P0}g_B \mathcal{L}_I L_\s{eff}-\alpha L\right)}.$ By measuring the output signal (at 75 kHz) with a lock-in amplifier and converting the voltage to optical power, we obtain the output signal power. The signal amplification is calculated as the difference between the output signal power and the input signal power.   

\begin{table}[htb]
\centering
\begin{tabular}{ll} 

Parameter description & Value\\
 \hline
 HCF attenuation (including the 0.5 dB \cod\ absorption) & 5.99 km$^{-1}$\\
 Signal power at the input of the HCF (inside the HCF)& -34 dBm\\
 Signal modulation frequency $f$ & 37.5 kHz\\
 Lock-in amplifier detection frequency & 75 kHz\\
 Lock-in amplifier, signal power (average optical power) to voltage transfer coefficient & 5.36 V/mW\\
 \hline
\end{tabular}
\caption{\textbf{Experimental details for signal amplification measurement.}}
\label{tab_amplificatio}
\end{table}

\section{Detailed experimental set-up for distributed temperature sensing}
\label{sec:dts_sensing}

\begin{figure*}[htbp]
  \centering{
  \includegraphics[width = 1\linewidth]{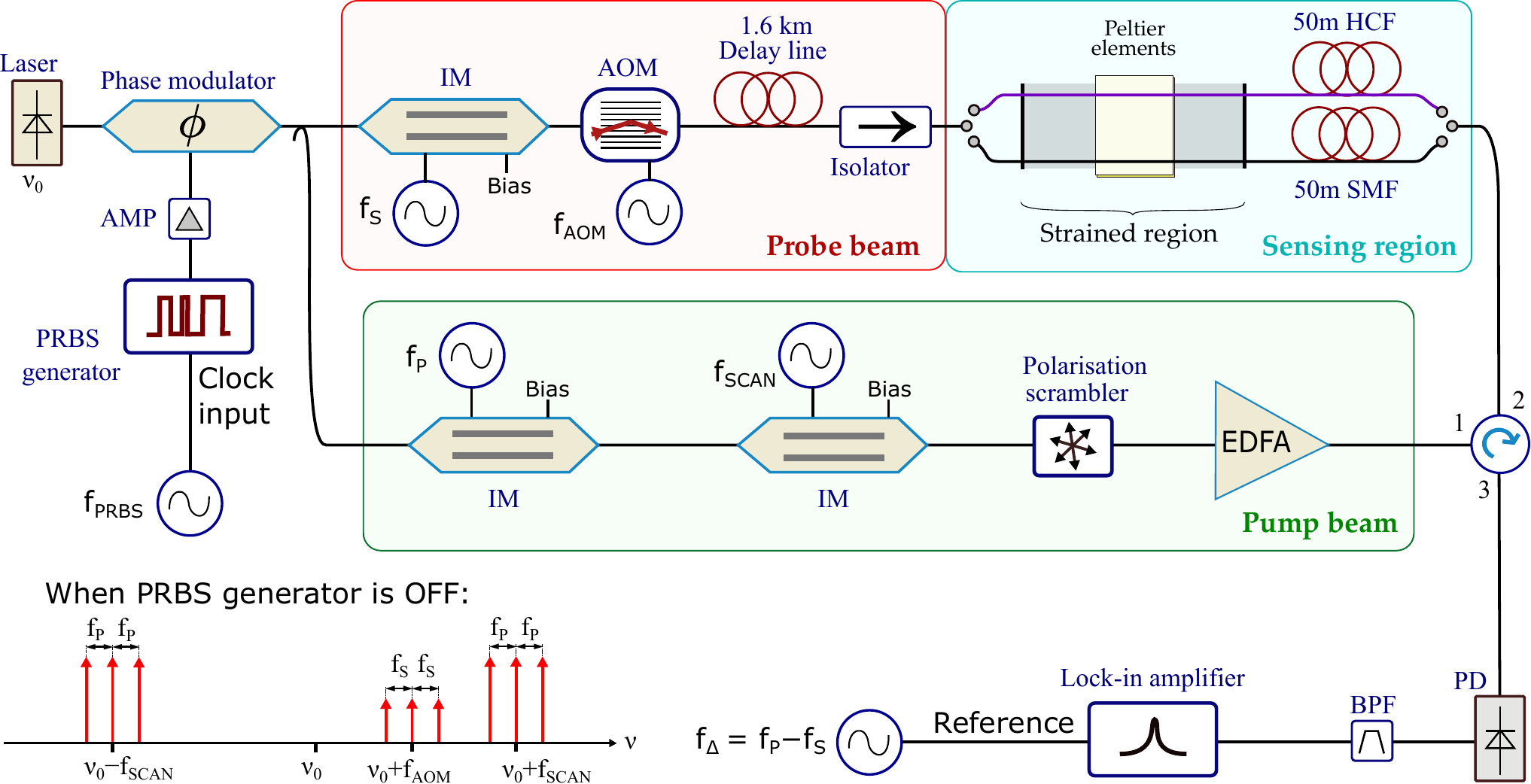}
  }
  \caption{\textbf{Experimental set-up for distributed temperature sensing in a gas-filled HCF by implementing a BOCDA.} The implementation is a combination of BOCDA technique with the dual intensity modulation method (see text). Note that the three radio-frequency sources generating the frequencies $f_P$, $f_S$ and $f_\Delta$ are synchronised to the same frequency standard. PRBS, pseudo-random bit sequence; AOM, acousto-optic modulator; IM, intensity modulator; EDFA, erbium-doped fibre amplifier; PD, photodetector; AMP, radio-frequency amplifier; BPF, band-pass filter. 
}
\label{fig:dts_schematic}
\end{figure*}

The experimental set-up we used is shown in Fig. \ref{fig:dts_schematic}. It is essentially a combination of a Brillouin optical correlation-domain analyser (BOCDA) based on phase modulation \cite{denisov2016going} with a dual-intensity modulation, presented in section (\ref{sec:dualintensitymodulation}), in order to filter out the pump reflection at the coupling interface. An intensity modulator generates two sidebands for the pump beam, used for the scanning. An AOM placed on the probe line shifts the probe frequency and thus breaks the pump sideband symmetry. Therefore, only one pump sideband is necessary for the scanning. The other unused sideband is not filtered but does not interfere in any way with the measurement. Polarisation is handled by placing a polarisation scrambler on the pump line. When the pseudo-random bit sequence (PRBS) generator is turned on, the random phase modulation applied to both pump and probe beams allows the acoustic waves to grow only in a precise location inside the fibre in which the phase of both pump and probe beams correlates and, thus, enables the experimental set-up to be used for distributed temperature sensing.

\subsection{Dual intensity modulation implementation}
The pump and probe are intensity-modulated with a RF modulation frequency of $f_P = 1$ MHz and $f_S =1.075$ MHz, respectively, so that the frequency difference is $f_\Delta = f_P - f_S = 75$ kHz. The modulator bias configuration was tuned to the quadrature point, as illustrated in the spectrum of Fig. \ref{fig:dts_schematic}.

\subsection{PRBS generator}
The PRBS generator reference clock $f_\s{PRBS}$ was set to 11.7 GHz for HCF and to 4.5 GHz for SMF, leading to a bit duration $\tau_b$ of 85.5 ps and 222.2 ps, respectively. These frequencies were experimentally chosen as the highest frequencies for which our system still gave a reasonable signal-to-noise ratio (SNR > 10). The resulting spatial resolutions, computed as the inverse of the bit duration, were 1.28 cm and 2.32 cm for HCF and SMF respectively. The exact spatial resolutions are slightly higher (i.e. smaller length) than these values \cite{Denisov_thesis_BOCDA}. The PRBS sequence length $N_\s{PRBS}$ was chosen as $N_\s{PRBS} = 2^{15}-1$ such that the sequence length (420 m and 760 m for HCF and SMF, respectively) is large enough to cover the whole fibre length.
\subsection{Phase modulator}
For proper functioning of the system, when the PRBS generator outputs a bit '1', a phase shift precisely equal to $\pi$ has to be reached as fast as possible to avoid unwanted activation of acoustic wave in the fibre at specific locations corresponding to the switching of the PRBS sequence \cite{Denisov_thesis_BOCDA}. Hence, a low $V_\pi$ ($V_\pi \approx 4 V$), high frequency ($20$ GHz) phase modulator was used in this experiment.

\subsection{Correlation location scanning}
In order to adjust the optical path length difference, an optical delay-line able to delay an optical signal up to a time $\tau = L\cdot n/c$, where $L$ is the measurement fibre length, would be required. Building such a delay-line is very challenging. Thus, another option is commonly adopted. This alternative method takes advantage of the fact that the total duration of the PRBS sequence, and thus the time at which the next sequences will start, depends on the bit duration. This technique consists of adding a fixed delay-line, i.e. an optical fibre of length $L_\s{dl} > N_\s{PRBS}\cdot\tau_b\cdot n_g/c$  (in our case, $L_\s{dl} = $1.6 km) in either the probe or pump path and slightly tuning the PRBS bit duration to shift the time at which the next sequences will start. The change of the correlation peak location $\Delta z_\s{cp}$ as a result of a slight change of the PRBS clock frequency $\Delta f_\s{PRBS}$ is found to be given by \cite{desmond_BOCDA}:
\begin{equation}
\Delta z_\s{cp} = \frac{1}{2}\frac{\Delta f_\s{PRBS}}{f_\s{PRBS}}\cdot L_\s{dl}.
\end{equation}
Thus, in our case, the total change of the PRBS clock frequency $\Delta f_\s{PRBS}$ required to scan the whole fibre is: $\Delta f_\s{PRBS}/f_\s{PRBS} =$ 6\%. This change of the PRBS clock frequency results in a 6\% change in the spatial resolution, which can be neglected here.

\subsection{Pump power}
The pump power, before entering either the HCF or the SMF, was set to 100 mW. In the case of the HCF, further increase in the power did not lead to an increase in the SNR. The power is estimated to be limited by the reflection of the amplified spontaneous emission noise from the erbium-doped fibre amplifier, directly entering into the photodetector.

\subsection{Probe signal acquisition}
Probe signal was recorded by using a standard detector ($\s{NEP} = 20~\s{pW/\sqrt{Hz}}$) followed by a 75 kHz bandpass RF filter to select only the signal of interest. Measurement spectra were recorded using a 7.8 Hz equivalent noise bandwidth (the lock-in amplifier was set to 10 ms time constant with 24 dB/octave filter slope) and their peak gain frequency were estimated using the quadratic fitting described in section \ref{sec:quadratic_fitting}.

\subsection{Scanning, quadratic fitting and repeatability estimation}
\label{sec:quadratic_fitting}

The scanning steps for HCF and SMF was 0.5 MHz and 1 MHz, respectively.

The quadratic fitting algorithm first applies a low-pass filter to the Brillouin gain data points and takes the maximum value in order to find the approximate peak position. Then, it keeps 17 original data points on each side from the approximate peak data point and discards the remaining data points. Note that these data points are the original data (without low-pass filtering). Finally, it performs a least-square quadratic fitting on these original data points.

The repeatability for the HCF and the SMF is computed as the average along 21 position points of the standard deviations of 8 measurements, previously fitted with the aforementioned quadratic fitting.

\subsection{Test bench}
\begin{figure}[htbp]
	\centering
	\includegraphics[width=\linewidth]{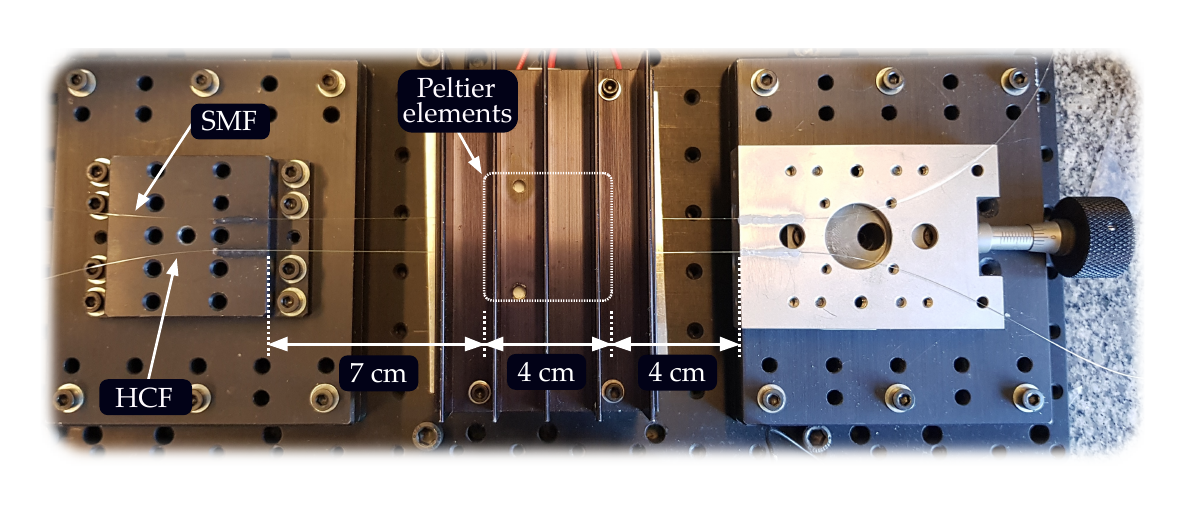}
	\caption{\textbf{Test bench for distributed temperature sensing.} Test bench to demonstrate the absence of strain sensitivity. A 4-cm-long temperature stage is placed in the middle of a 15-cm-long strain stage, allowing to apply strain and temperature changes in the same region of the fibres. Both SMF and HCF cross the test bench parallel to each other and are placed close to each other for a fair comparison.}
	\label{fig:dts_testbench}
\end{figure}

In order to demonstrate the absence of strain sensitivity and to perform a fair comparison with the standard single-mode fibre, a dedicated test bench was built, which enables us to apply both strain and temperature changes at the same fibre location. The ends of both our 50-m-long HCF and of a 50-m-long ITU G.652 single-mode fibre used for comparison are placed inside this test bench, which consists of two parts:
\begin{itemize}
	\item {Temperature stage.} The fibres are "sandwiched" between two 4-cm-long Peltier elements, placed below and above the two fibres, respectively. The thermal conductivity between the Peltier elements and the fibres is ensured by the presence of thermal paste. Furthermore, two metallic radiators are placed on the other side of each Peltier element to provide/dissipate heat from/to the environment. In addition, a fan forces the flow of air through these radiators in order to ensure a sharp temperature transition.
	
	\item {Strain stage.} The previously described temperature stage is placed in the middle of a 15-cm-long strain stage. On one side, the fibres are glued onto a fixed metallic plate while on the other side, they are glued onto a displacement stage. Note that the coating of the two fibres was removed at the gluing points.
\end{itemize}
A picture of this test bench is provided in Fig. \ref{fig:dts_testbench}, including the various lengths. Note that the fan is not visible in the picture.

\section{Theoretical calculation of the Brillouin gain}

\begin{table}[h]
\centering
\begin{tabular}{lll} 
Parameter name & Parameter description & Value (for \cod\ at 41 bar)\\
 \hline
 $A^{\rm ao}_{\rm eff}$ & Acousto-optic overlap effective area & 80 $\rm {\upmu m^2}$ \\
 $\eta_s$ & Shear viscosity \cite{wang2019bulk} & $1.5\times 10^{-5} $ Pa$\cdot$s \\ 
 $\eta_b$ & Bulk viscosity \cite{pan_power_2005} & $4\times 10^{-6}$ Pa$\cdot$s \\ 
 $\kappa$ & Thermal conductivity & 0.01662 $\rm {W\cdot m^{-1}K^{-1}}$\\
 $C_P$ & Specific heat at constant pressure & 846 $\rm {J\cdot kg^{-1}K^{-1}}$\\
 $\rho$ & Gas density & 72.77 kg/$\rm {m^3}$\\
 $\gamma$ & Heat capacity ratio (adiabatic index) & 1.3\\
 $n$ & Refractive index of the gas & 1.01804\\
 $\rm T$ & Temperature & 298 K\\
 $v_a$ & Acoustic velocity & 243.6 m/s\\
 \end{tabular}
 \caption{\textbf{Detailed parameters for the theoretical calculation of the Brillouin gain.} Note that ideal gas model was assumed for the calculation of $C_P$, $\rho$ and $\gamma$. More complex calculations taking into account non-ideal gas models do not significantly alter the values.}
 \label{tab_theoretical_calculation_gain}
\end{table}

\noindent
The acousto-optic overlap effective area is given by \cite{kobyakov2010stimulated}:
\begin{equation}
A^{\rm ao}_{\rm eff} = \left[\frac{\langle f^2(x,y)\rangle}{\langle\xi(x,y) f^2(x,y)\rangle}\right]^2 \langle\xi ^2(x,y)\rangle,
\end{equation}
\noindent
where $f^2(x,y)$ and $\xi(x,y)$ are the transverse optical intensity profile and acoustic pressure profile of the fibre, respectively, and where the operator $\langle ..\rangle$ performs an integration over the entire fibre cross-section. We use the numerical simulation results shown in Figs. 1(c) and (d) in the main manuscript and make overlap integration to obtain the acousto-optic overlap effective area: $A^{\rm ao}_{\rm eff} =80~\s{\upmu m^2}$. By plugging all the parameters of 41-bar \cod\ into Eq. (1) in the main manuscript, we can calculate the theoretical Brillouin gain to be 1.86 $\rm {m^{-1}W^{-1}}$, which is very close to the measured Brillouin gain 1.68 $\rm {m^{-1}W^{-1}}$. By inserting all the parameters into Eq. (6) in the main manuscript, we obtain a Brillouin linewidth of 4.3 MHz, which is also close to the measured linewidth 3.65 MHz.   

\section{Calculation of Raman gain coefficient}

\begin{figure*}[htbp]
  \centering{
  \includegraphics[width = 0.8 \linewidth]{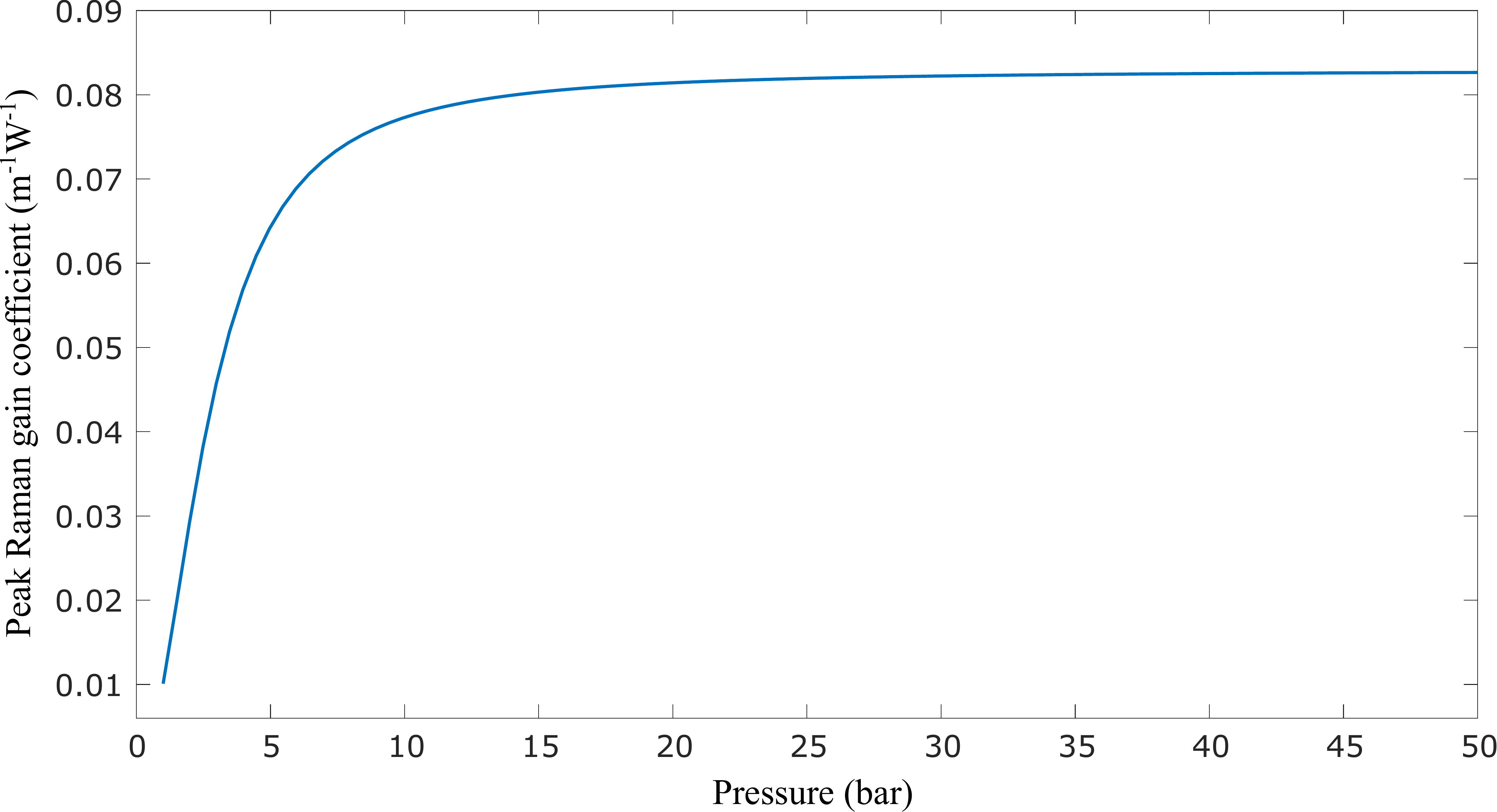}
  }
  \caption{\textbf{Raman gain coefficient as a function of hydrogen pressure at a wavelength of 1.55 $\rm {\mathbf{\upmu m}}$ and room temperature (Q(1) vibrational transition).} 
}
\label{fig:RamanGain}
\end{figure*}

\noindent
So far, hydrogen gas shows the highest Raman gain (at a detuning frequency of 125 THz for the Q(1) vibrational transition) of any gas \cite{mridha_thresholdless_2019}. The peak plane-wave steady-state Raman-gain coefficient, $g_R$ (in units of cm/W), for the Q(1) transition for pump-laser wavelengths from 190 nm to 2 $\rm {\upmu{m}}$, densities of 1-100 amagats, at room temperature (298 K) is given as \cite{bischel_wavelength_1986,russell_hollow-core_2014}:
\begin{equation}
\label{Eq:Raman_gain}
{g_R}= 9.37\times{10^6}\cdot\frac{(52\rho/\Delta {\nu})(\nu {_p}-4155)}{{(7.19\times{10^9}-{\nu {_p}}^2)^{2}}},
\end{equation}
where $\rho$ is the density in amagats, $\Delta \nu$ is the Raman linewidth in MHz, given by $\Delta \nu=(309/\rho)+51.8\rho$ at room temperature, $\nu _p$ is the pump laser frequency in inverse centimetres. This means when the density is above 9 amagats, $\Delta \nu \approx 51.8\rho$ MHz. At room temperature, a pressure of 1 amagat corresponds to 1.1 bar. As a result, the peak Raman gain is independent of density (i.e it is saturated) when the pressure is above 10 bar because the Raman linewidth is proportional to the pressure (pressure broadening region, caused by the onset of inelastic rotational collisions \cite{bischel1986temperature}). By substituting this pressure into Eq. \ref{Eq:Raman_gain}, we can compute the Raman gain (in units of cm/W) as a function of pressure. The saturated maximum Raman gain coefficient is calculated as $4.2\times {10^{-12}}$ m/W at 1.55 $\rm {\upmu m}$. Assuming the use of the same HCF with a core diameter of 10.9 $\rm {\upmu m}$ and an optical effective mode field area of 51 $\rm {\upmu m}^2$ (calculated with COMSOL), the peak Raman gain coefficient (in units of $\rm m^{-1}W^{-1}$) as a function of pressure is plotted in Fig. \ref{fig:RamanGain}. The highest Raman gain for more than 10 bar (e.g. 41 bar) hydrogen is 0.08 $\rm m^{-1}W^{-1}$, which is more than 20 times smaller than the Brillouin gain in gas demonstrated in this paper.

\section{Acoustic velocity in \cod\ at different gas pressures}
\begin{figure*}[htbp]
  \centering{
  \includegraphics[width = 0.8 \linewidth]{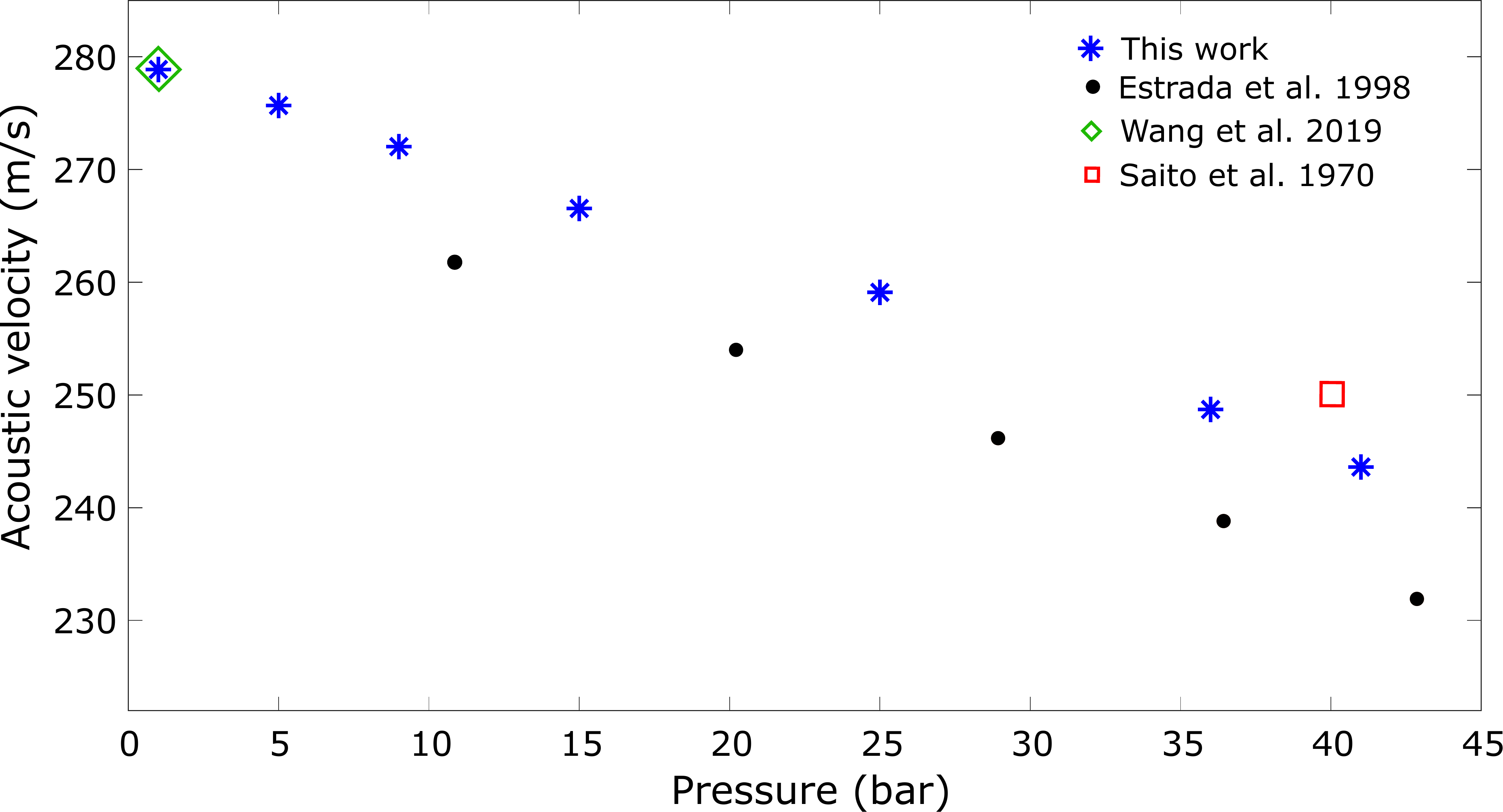}
	\caption{\textbf{Acoustic velocity in \cod\ at different gas pressures.}}
}
\label{fig:acoustic_velocity_pressure}
\end{figure*}

\noindent
We measured the Brillouin gain spectrum at different gas pressures. The pump-probe frequency detuning at the peak gain is called Brillouin frequency shift and is given by \cite{wang2019bulk}: 
\begin{equation}
\label{Eq:acoustic_velocity}
\nu_B = \frac{2n_\s{eff}v_a}{\lambda_P},
\end{equation}
where $n_\s{eff}$ is the effective refractive index of the optical mode, $v_a$ is the acoustic velocity and $\lambda_P$ is the pump wavelength. As an approximation, we used the gas refractive index $n$ as the effective refractive index $n_\s{eff}$. We can derive the measured acoustic velocity (shown by blue stars in Fig. \ref{fig:acoustic_velocity_pressure}) from the Brillouin frequency shift by using Eq. \ref{Eq:acoustic_velocity}. The black dots in Fig. \ref{fig:acoustic_velocity_pressure} is the measured results for low frequency acoustic waves (several kilohertz) from Ref. \cite{estrada1998speed}. Our results show a very similar trend. The mismatch of the absolute value is probably due to the different values of the bulk modulus at different acoustic frequencies (our frequency range is $\sim 320$ MHz), possibly caused by translational-vibrational relaxation processes, since \cod\ is a polyatomic gas. The green diamond in Fig. \ref{fig:acoustic_velocity_pressure} represents the measured acoustic velocity for a 500 MHz acoustic wave at 1 bar from Ref. \cite{wang2019bulk}. This value matches well with our measurement. 
The red square in Fig. \ref{fig:acoustic_velocity_pressure} shows the measured acoustic velocity at 40 $^{\circ}$C from Fig. 5 in Ref \cite{saito_measurement_1970}. This value is a little bit higher than our result, probably because their temperature is higher than in our experiments.

\section{Axial strain finite-element simulation}
An axial strain applied on the fibre will lead to the following effects:

\begin{itemize}
	\item Due to the Poisson effect, the honey-comb structure will be distorted, modifying the effective refractive index of the optical mode.
	\item In particular, the holes will shrink and the fibre will elongate, leading to a change in the volume available for the gas along the fibre.
	\item The silica refractive index will change due to the photo-elastic effect, leading to a change in the effective refractive index of the optical mode.
\end{itemize}

\begin{figure*}[htbp]
  \centering{
  \includegraphics[width = 1 \linewidth]{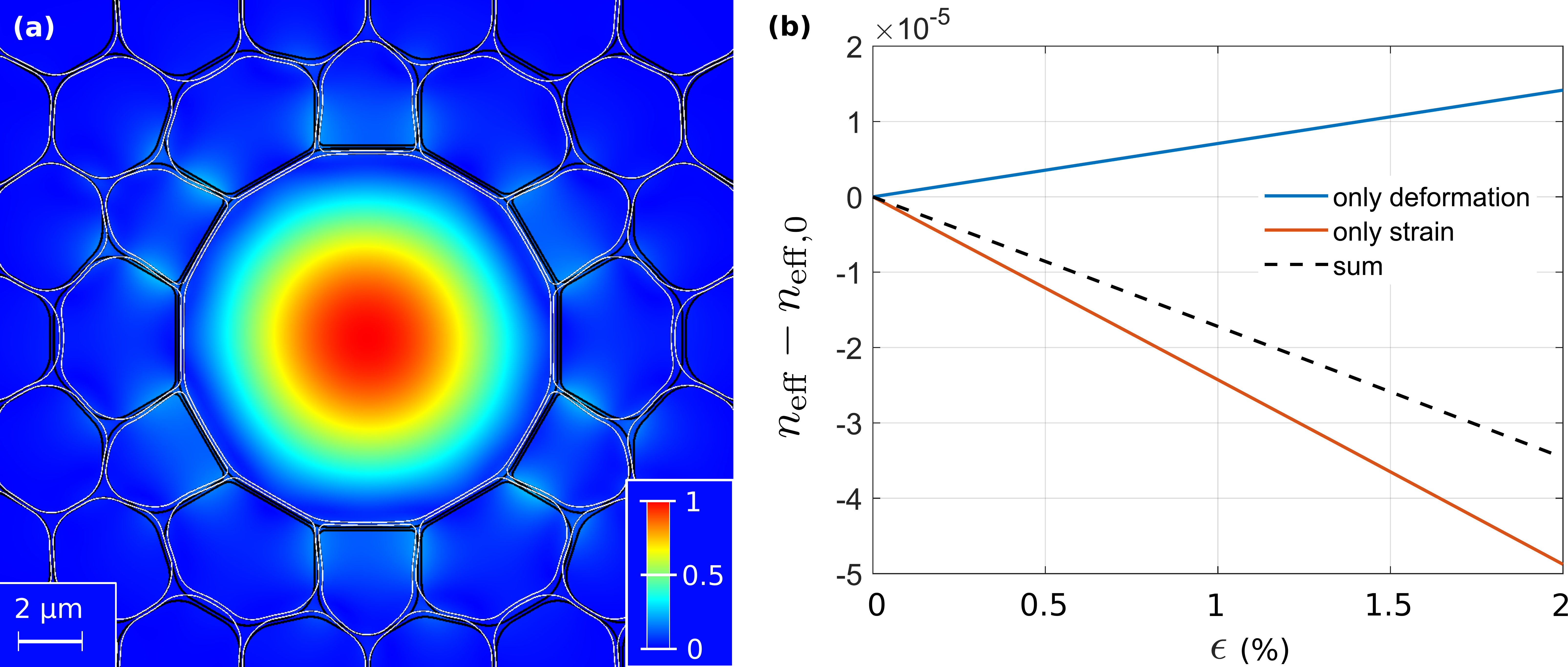}
	\caption{\textbf{Simulation of the deformation of the HCF's cross-section in response to an axial strain.} \textbf{(a)} Normalised electric field magnitude of one of the two fundamental modes (optical wavelength: 1.55 $\upmu$m) in the case of a hypothetical axial strain of 20\%. White lines draw the deformed structure while black lines draw the original, undeformed structure. \textbf{(b)} Evolution of the effective refractive index, $n_\s{eff}$, as a function of the strain and relative to the effective refractive index in absence of strain, $n_\s{eff,0}$. The blue line accounts only for the structure deformation, the red line accounts only for the photo-elastic effect and the dotted black line accounts for the total evolution (i.e. the sum of these two contributions).}
}
\label{fig:dts_strain_neff}
\end{figure*}

In order to quantify these effects, a finite-element simulation has been performed. To this end, deformations of a 3D slice of the HCF subject to an axial strain, $\epsilon$, were computed. The deformations in the cross-section plane have subsequently been used to compute the resulting change in the effective refractive index $n_\s{eff}$. Although this computation was performed assuming 40 bar gas pressure in the holes, the results only weakly depend on the gas pressure. Figure \ref{fig:dts_strain_neff}(a) shows the deformed geometry (white lines) compared to the original geometry (black lines) for a hypothetical strain of 20\%, as well as the normalised electric field for one fundamental mode in the deformed structure. Figure \ref{fig:dts_strain_neff}(b) shows the evolution of the effective refractive index, $n_\s{eff}$, relative to the the effective refractive index in the undeformed case, $n_\s{eff, 0}$, as a function of strain, from 0 up to 2\%. The red trace shows the contribution of the photo-elastic effect while the blue trace shows the contribution of the structure deformation (as shown in Fig. \ref{fig:dts_strain_neff}(a)). These two contributions show a linear evolution equal to: $\partial n_\s{eff}/\partial\epsilon=-2.4\times 10^{-3}$ and $\partial n_\s{eff}/\partial\epsilon=+7.08\times 10^{-4}$ for the photo-elastic effect and for the deformation, respectively. We can see that the photo-elastic effect dominates and is partly compensated by the contribution of the deformation. The black dotted line shows the total evolution of $n_\s{eff}$ as a function of the applied axial strain and has a linear value of: $n_\s{eff}/\partial\epsilon=-1.7\times 10^{-3}$.

We are now in position to estimate the local change in the Brillouin frequency shift due to an axial strain along the HCF. The change of the Brillouin frequency shift due to an axial strain $\epsilon$ can be written as:
\begin{equation}
\frac{\dd\nu_B}{\dd \epsilon} = \frac{\nu_B}{n}\frac{\partial n}{\partial \epsilon} + \frac{\nu_B}{V_a}\frac{\partial V_a}{\partial \epsilon}.
\end{equation}
Considering the 40 bar CO$_2$ Brillouin frequency shift of 320 MHz and $n\approx 1$, the first term can be directly evaluated to:
\begin{equation}
\frac{\nu_B}{n}\frac{\partial n}{\partial \epsilon} = -544~\s{mHz/\upmu\upepsilon}.
\end{equation}
In order to evaluate the second term, we consider that the fibre, of length $L$, is axially strained over a length $l$. In these conditions, the total change in the gas pressure $P$ is:
\begin{equation}
\frac{\Delta P}{P}=-\frac{\Delta V}{V} = -\epsilon\left(1-2\kappa\right)\frac{l}{L},
\end{equation}
where we considered the change in volume of a cylinder of volume $V_c$, stretched by a strain $\epsilon_c$, that can be expressed as: $\Delta V_c/V_c = \epsilon_c\left(1-2\kappa\right)$, with $\kappa$ being the Poisson ratio. Taking as an extreme example a 40-bar gas-filled 50-m-long fibre whose entire length is strained by an axial strain of 2\% (i.e. 1 meter elongation), the total pressure change would be: $\Delta P = -0.5$ bar. As the pressure-dependent Brillouin frequency was found to be of $\approx$ $-$1 MHz/bar, the total change in the Brillouin frequency due to the gas pressure variation would be of $+0.5$  MHz. In addition, the change in the Brillouin frequency due to the effective refractive index variation would be of $-10.8$ kHz. We can see that these two effects have an opposite sign and partly compensate for each other. For example, if we consider the same example but with the strain applied along a 20 centimeter-long portion of the fibre only, the change in the Brillouin frequency due to the gas pressure variation would be $10$ kHz and would almost perfectly compensate for the change in the Brillouin frequency due to the effective refractive index variation. In summary, we demonstrated that the change in the Brillouin frequency due to an applied axial strain can be neglected in normal operation.

\section{Response of a Mach-Zehnder intensity modulator}

\label{appendix_intensity_modulation}

\subsection{Introduction}
\begin{figure}[htbp]
	\centering
	\includegraphics[width=\linewidth]{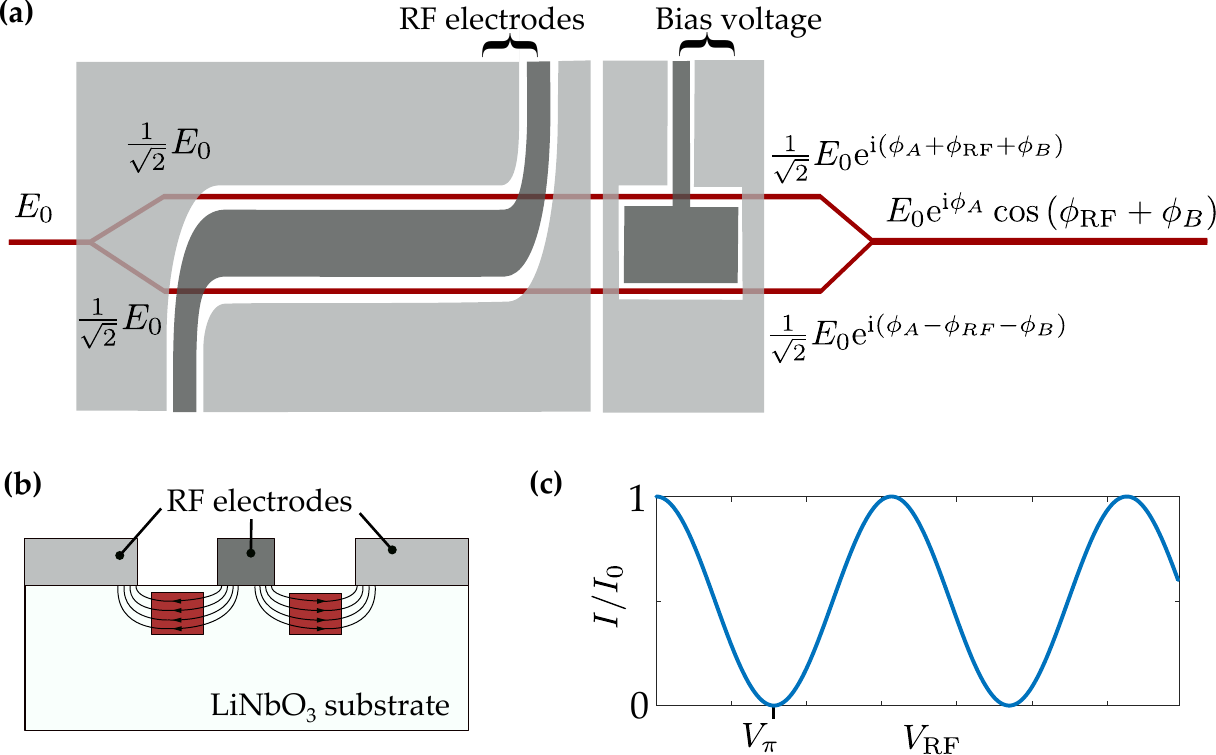}
	\caption{\textbf{Mach-Zehnder modulator.} \textbf{(a)} Top view of a typical Mach-Zehnder modulator. The waveguides are drawn in red and the electrodes in gray (dark gray for the signal electrode and light gray for the ground electrode). Light propagates from left to right. Expressions indicate the electric field at various positions. \textbf{(b)} Cross-section of a typical  Mach-Zehnder modulator showing the lithium niobate ($\s{LiNbO_3}$) substrate, the waveguide of the two arms in red and the electrodes in gray (dark gray for the signal electrode and light gray for ground electrode). The arrows show the orientation of the electric field crossing the waveguides when the RF voltage is positive. \textbf{(c)} Normalised output light intensity as a function of the RF voltage, $V_\s{RF}$, following a cosine function.}
	\label{fig:machzehndermodulator}
\end{figure}

In this section, we discuss the Mach-Zehnder modulator and derive an expression for the output intensity in the special case of a RF sinusoidal modulation. A typical Mach-Zehnder modulator consists of a lithium niobate ($\s{LiNbO_3}$) substrate in which optical waveguides are imprinted. Metallic electrodes are subsequently deposited on top of the substrate. A top view and a cross-section of a typical Mach-Zehnder modulator is shown in Fig. \ref{fig:machzehndermodulator}(a) and (b) respectively \cite{ixblue_bias}. 

The input waveguide is split into two arms and the top electrodes are used to induce an electric field into the two arms. Thanks to Pockels effect, the induced electric field leads to a slight increase in the refractive index to one arm and a slight decrease in the other arm. The two waveguides are then merged into one output waveguide again, resulting in an interference of the two optical fields. The extinction ratio is a parameter measuring the intensity ratio between constructive interference and destructive interference and its value is typically $\approx$ 25 dB (however, in the modulators used in this work, it was larger than 35 dB). Figure \ref{fig:machzehndermodulator}(c) shows the normalised output intensity (neglecting insertion loss) as a function of the applied RF voltage, following a cosine variation. The voltage change required to go from a constructive interference to a destructive interference is called $V_\pi$.

In order to simplify the calculation, we assume a perfect modulator as possessing the following characteristics:
\begin{itemize}
	\item The modulator is lossless. However, insertion loss can be easily added by simple multiplication of the results with a loss factor.
	\item The modulator has an infinite bandwidth. This essentially means that the modulation frequency is much smaller than the modulator cut-off bandwidth and simplifies the derivation by ignoring the effects of a finite bandwidth.
	\item The modulator has exactly equal arm lengths. While never the case in reality, compensation for unequal arm lengths and temperature drifts can be easily done by adjusting the modulator's bias voltage.
	\item In the case of a destructive interference, no light intensity is present at the output (infinite extinction ratio).
\end{itemize}

As shown in Fig. \ref{fig:machzehndermodulator}(a), in presence of an input electric field $E_0$, the electric field at the output of the modulator reads:
\begin{equation}
E = E_0\ex{\phi_A}\cos{\left(\phi_\s{RF}+\phi_B\right)},
\label{machzehndermodulatorelectricfieldgeneralequation}
\end{equation}
where $\ex{\phi_A}$ is the phase shift induced by the light propagation across the modulator, $\phi_\s{RF} = \pi V_\s{RF}/2V_\pi$ is the phase shift induced when applying a RF voltage $V_\s{RF}(t)$ to the electrodes and $\phi_B$ is the bias phase shift induced when applying a certain voltage to the bias port. Since $\ex{\phi_A}$ is a common phase shift that is simply due to transmission across the device, we disregard it from now on.

We now apply a sinusoidal RF voltage of frequency $\Omega/2\pi$ and of amplitude $V$:
\begin{equation}
V_\s{RF}(t) = V\sin{(\Omega t)}.
\end{equation}
The RF phase shift is:
\begin{equation}
\phi_\s{RF} = \frac{\pi}{2}\frac{V_\s{RF}(t)}{V_\pi}=\varsigma\sin{(\Omega t)},
\label{machzehnderrfsinusoidalvoltage}
\end{equation}

\noindent
where $\varsigma = \pi V/2V_\pi$ is called the modulation depth. Inserting (\ref{machzehnderrfsinusoidalvoltage}) into (\ref{machzehndermodulatorelectricfieldgeneralequation}) and applying Jacobi-Anger's identity, we obtain:
\begin{equation}
\begin{split}
E &= E_0\Re{\left\{\ex{\i\varsigma\sin{\Omega t}}\ex{\i\phi_B}\right\}} = E_0\Re{\left\{\ex{\i\phi_B}\sum_{n}{J_n(\varsigma)\ex{\i n\Omega t}}\right\}}\\
&= E_0\sum_{n}{J_n(\varsigma)\cos{\left(n\Omega t + \phi_B\right)}},
\end{split}
\label{eqn_intensitymodfield}
\end{equation}

\noindent
where $J_n$ are the Bessel functions of the first kind and $n$ is a scalar going from $-\infty$ to $+\infty$. We will use this equation to derive an expression for the output intensity, as observed when connecting the modulator output to a photodetector.

\subsection{Intensity at a photodetector}

We now calculate the output intensity $I = \norm{E}^2/2\eta$, where $\eta$ is the medium impedance. We define $I_0= \norm{E_0}^2/2\eta$ as the input light intensity and take the magnitude squared of Eq. (\ref{eqn_intensitymodfield}):
\begin{equation}
\begin{split}
I(t) =&I_0\sum_{n}\sum_{m}J_n(\varsigma)J_m(\varsigma)\cos{\left(n\Omega t + \phi_B\right)}\cos{\left(m \Omega t+\phi_B\right)}\\
=\frac{1}{2}&I_0\sum_{n}\sum_{m}J_n(\varsigma)J_m(\varsigma)\big[\cos{\bigl((n+m)\Omega t + 2\phi_B\bigr)}+\cos{\bigl((n-m) \Omega t\bigr)}\big].
\end{split}
\end{equation}
At this point, it is useful to rearrange the terms by defining the following variable change: $p = n + m$ and $q = n - m$. As illustrated in table \ref{table:sumrearrange}, this procedure is equivalent to switching from a horizontal/vertical indices scanning to a diagonal one. Note that $p$ and $q$ should have the same parity: when $p$ is even, $q$ also has to be even and when $p$ is odd, $q$ also has to be odd. We thus separate the sum into two parts; one for odd values of $p$ and $q$ and one for even values of $p$ and $q$.
\begin{table*}[htbp]
	\caption{Illustration of the variables change: $p = n+m$ and $q=n-m$. Original horizontal and vertical scanning for $n$ and $m$ indices of the summation is changed to a diagonal scanning for $p$ and $q$. Note that although the indices shown run from $-3$ to $3$, the actual sum is infinite.}
	\includegraphics[width=\linewidth]{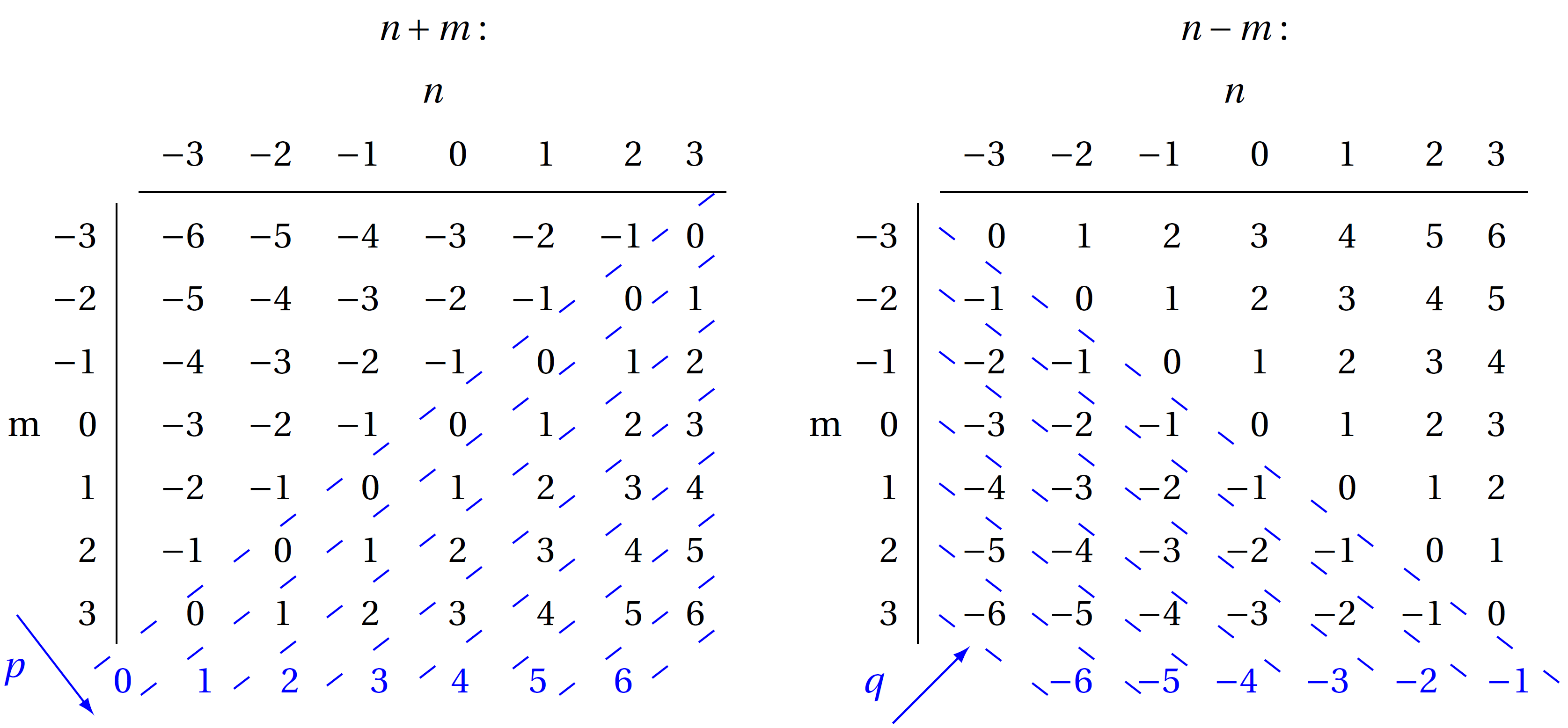}
	\label{table:sumrearrange}
\end{table*}
The sums are thus rearranged and written as:
\begin{equation}
\begin{split}
I(t)=&\frac{1}{2}I_0\sum_{p}\sum_{q}J_{p+q}(\varsigma)J_{p-q}(\varsigma)\big[\cos{\left(2p\Omega t+2\phi_B\right)}+\cos{\left(2q\Omega t\right)}\big]\\ +&\frac{1}{2}I_0\sum_{p}\sum_{q}J_{p+q+1}(\varsigma)J_{p-q}(\varsigma)\big[\cos{\left((2p+1)\Omega t+2\phi_B\right)}+\cos{\left((2q+1)\Omega t\right)}\big].
\end{split}
\label{eq:rearrangedsums}
\end{equation}
We then use the following identities:
\begin{equation}
\begin{split}
&\sum_{q}J_{p+q}(\varsigma)J_{p-q}(\varsigma) = J_{2p}(2\varsigma)\\
&\sum_{p}J_{p+q}(\varsigma)J_{p-q}(\varsigma) = \begin{cases}1,\quad q = 0\\0, \quad\mathrm{otherwise}\end{cases}\\
&\sum_{q}J_{p+q+1}(\varsigma)J_{p-q}(\varsigma) = J_{2p+1}(2\varsigma)\\
&\sum_{p}J_{p+q+1}(\varsigma)J_{p-q}(\varsigma) = 0,\\
\end{split}
\end{equation}
which enable rewriting Eq. (\ref{eq:rearrangedsums}) as:
\begin{equation}
I(t)=\frac{1}{2}I_0\Bigl(1+\sum_{n}{J_{n}(2\varsigma)\cos{\left(n\Omega t + 2\phi_B\right)}}\Bigr).
\end{equation}
In order to isolate the different frequency components, we can modify the result as follows:
\begin{equation}
\begin{split}
I(t)=\frac{1}{2}I_0\Bigl(1+J_0(2\varsigma)\cos{\left(2\phi_B\right)} &+ \sum_{n=1}^{\infty}J_n(2\varsigma)\cos{\left(n\Omega t+2\phi_B\right)}\\
&+ \sum_{n=1}^{\infty}\left(-1\right)^n J_n(2\varsigma)\cos{\left(n\Omega t -2\phi_B\right)}\Bigr),\\
\end{split}
\end{equation}
where the relation $J_{-n}(\varsigma) = \left(-1\right)^n J_{n}(\varsigma)$ has been used. Using trigonometric relations and again separating odd and even frequencies, it follows that:
\begin{equation}
\begin{split}
I(t)=I_0\Bigl(&\frac{1}{2}+\frac{J_0(2\varsigma)}{2}\cos{\left(2\phi_B\right)}\\
&-\sin{\left(2\phi_B\right)}\sum_{n=1}^{\infty}J_{2n-1}(2\varsigma)\sin{\bigl((2n-1)\Omega t\bigr)}\\
&+\cos{\left(2\phi_B\right)}\sum_{n=1}^{\infty}J_{2n}(2\varsigma)\cos{\left(2n\Omega t\right)}\Bigr),
\end{split}
\end{equation}
where the first line represents the DC part, the second line represents the odd harmonics and the third line represents the even harmonics. From this expression, four particular values of the bias $\phi_B$ can be distinguished:
\begin{itemize}
	\vspace{-\topsep}
	\item When $\phi_B=\pi/2 + z\pi,\quad z\in\Z$, only the even harmonics are present and the DC value is low. This configuration is usually referred to as "carrier-suppressed".
	\item When $\phi_B=0 + z\pi,\quad z\in\Z$, only the even harmonics are present and the DC value is high. This configuration is usually referred to as "full-carrier".
	\item When $\phi_B=\pi/4 + z\pi,\quad z\in\Z$, only the odd harmonics are present. This configuration is usually referred to as "quadrature point".
	\item When $\phi_B=3\pi/4 + z\pi,\quad z\in\Z$, only the odd harmonics are present. This configuration is also usually referred to as "quadrature point". The only difference with respect to the previous case is the presence of a $\pi$-phase shift for the output intensity modulation.
\end{itemize}

\subsection{Analysis of the three bias configurations}

\begin{figure}[htbp]
	\centering
	\includegraphics[width=\linewidth]{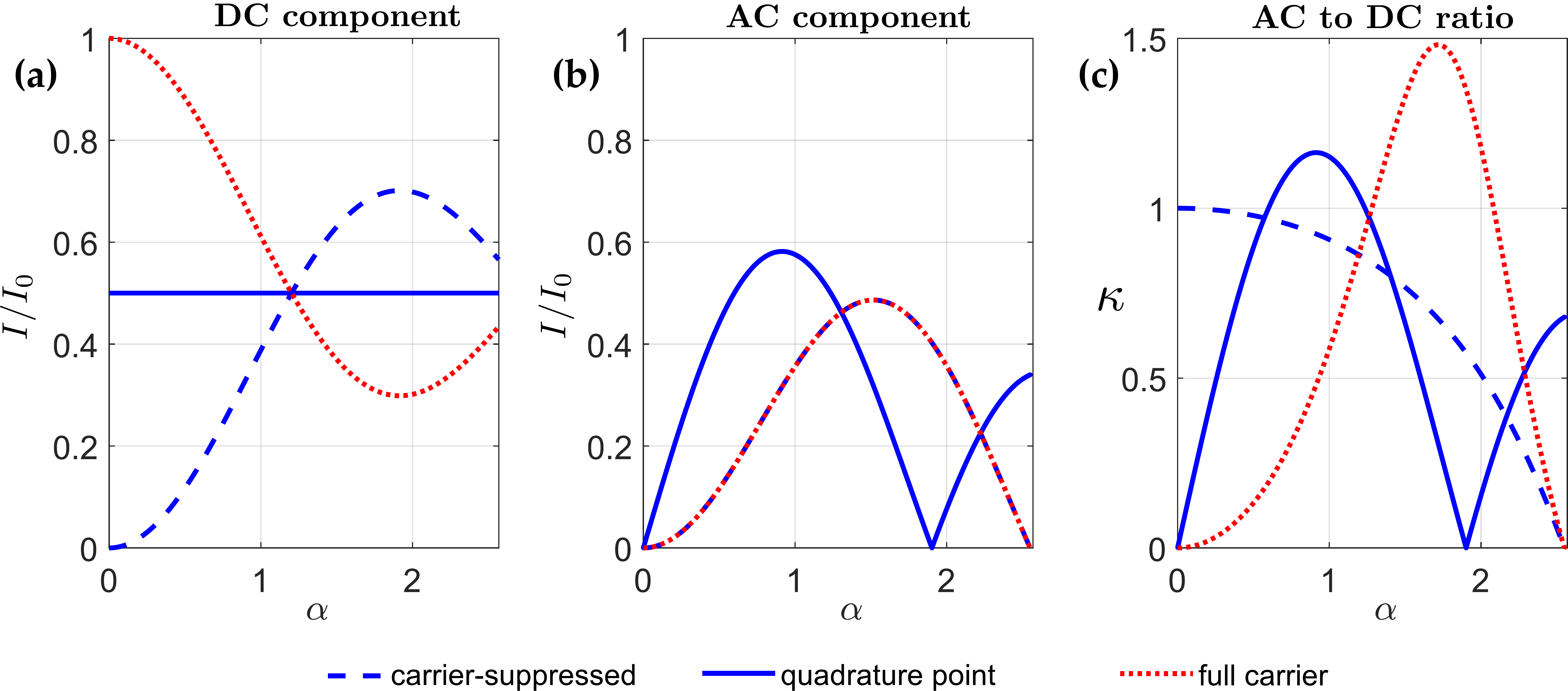}
	\caption{\textbf{DC and AC intensities.} \textbf{(a)} Evolution of the DC component of the modulator output intensity as a function of the modulation depth $\varsigma$. \textbf{(b)} Evolution of the AC component of the modulator output intensity as a function of the modulation depth $\varsigma$. \textbf{(c)} Evolution of the AC to DC ratio as a function of the modulation depth $\varsigma$.}
	\label{fig:kappa_plot}
\end{figure}

\subsubsection{Carrier-suppressed configuration}
When the bias is set so that $\phi_B=\pi/2 + z\pi,z\in\Z$, the output intensity contains only even multiples of the modulation frequency:
\begin{equation}
I(t)=I_0\Bigl(\frac{1}{2}-\frac{J_0(2\varsigma)}{2}-\sum_{n=1}^{\infty}J_{2n}(2\varsigma)\cos{\left(2n\Omega t\right)}\Bigr).
\end{equation}
When the modulation depth $\varsigma$ is moderate (< 2), the sum can be approximated by keeping only the first term:
\begin{equation}
I(t)\cong I_{\s{det}}\Bigl(1-\kappa_S\cos{\left(2\Omega t\right)}\Bigr),
\end{equation}
where $I_{\s{det}} = I_0(1-J_0(2\varsigma))/2$ is the time-averaged detected intensity and $\kappa_S=2J_2(2\varsigma)/(1-J_0(2\varsigma))$ is the ratio of the intensity at frequency $2\Omega$ to the DC intensity. The evolution of this parameter as a function of the modulation depth is plotted in Fig. \ref{fig:kappa_plot}(c).

\subsubsection{Quadrature point configuration}
When the bias is set so that  $\phi_B=\pi/4 + z\pi,z\in\Z$ or $\phi_B=3\pi/4 + z\pi,z\in\Z$, the output intensity contains only odd multiples of the modulation frequency:
\begin{equation}
I(t)=\frac{1}{2}I_0\Bigl(1\pm 2\sum_{n=1}^{\infty}J_{2n-1}(2\varsigma)\sin{\left(\left(2n-1\right)\Omega t\right)}\Bigr),
\end{equation}
where the sign in front of the sum depends on the choice of the bias points amongst the two cited above. When the modulation depth $\varsigma$ is moderate (< 2), the sum can be approximated by keeping only the first term:
\begin{equation}
I(t)\cong I_{\s{det}}\Bigl(1-\kappa_Q\sin{\left(\Omega t\right)}\Bigr),
\end{equation}
where this time $I_{\s{det}} = I_0/2$, which is the time-averaged detected intensity, is independent from the modulation depth. $\kappa_Q = 2J_1(2\varsigma)$ is the ratio of the intensity at frequency $\Omega$ to the DC intensity. The evolution of this parameter as a function of the modulation depth is plotted in Fig. \ref{fig:kappa_plot}(c).

\subsubsection{Full carrier configuration}
When the bias is set so that  $\phi_B=0 + z\pi,z\in\Z$, the situation is similar to the carrier-extinct situation except that the DC component is higher and the modulation has a phase shift:
\begin{equation}
I(t)=I_0\Bigl(\frac{1}{2}+\frac{J_0(2\varsigma)}{2}+\sum_{n=1}^{\infty}J_{2n}(2\varsigma)\cos{\left(2n\Omega t\right)}\Bigr).
\end{equation}
When the modulation depth $\varsigma$ is moderate (< 2), the sum can be approximated by keeping only the first term:
\begin{equation}
I(t)\cong I_{\s{det}}\Bigl(1+\kappa_F\cos{\left(2\Omega t\right)}\Bigr),
\end{equation}
where $I_{\s{det}} = I_0(1+J_0(2\varsigma))/2$ is the time-averaged detected intensity and $\kappa_F=2J_2(2\varsigma)/(1+J_0(2\varsigma))$ is the ratio of the intensity at frequency $2\Omega$ to the DC intensity. The evolution of this parameter as a function of the modulation depth is plotted in Fig. \ref{fig:kappa_plot}(c).

\subsubsection{Output intensity for the different carrier configurations}

We now use the aforementioned expressions to compare the output intensities for each bias configuration. Figure \ref{fig:kappa_plot}(a) shows the evolution of the DC component of the Mach-Zehnder output intensity as a function of the modulation depth $\varsigma$. Figure \ref{fig:kappa_plot}(b) shows the evolution of the AC component (at frequency $\Omega$ for the quadrature point, at frequency $2\Omega$ for the full-carrier and carrier-extinct configurations) of the Mach-Zehnder output intensity as a function of the modulation depth $\varsigma$ and Figure \ref{fig:kappa_plot}(c) plots the ratio of these two quantities (i.e. DC-to-AC intensity ratio), $\kappa$. It can be seen that the highest AC intensity among the three bias configuration is reached by the quadrature point. Moreover, for this bias configuration, the AC component peak appears at lower values of the modulation depth $\varsigma$. In practice, the RF source output power is typically limited to $\approx 25$ dBm and the $V_\pi$ of a standard modulator reaches 7 V, so the modulation depth can typically reach a maximum value of $\varsigma \approx 1.2$. Hence, the quadrature point bias configuration is ideal to achieve the highest modulation intensity. The parameter $\kappa$ is experimentally useful to quickly estimate the optical power at the modulation frequency by simply measuring the DC optical power using a power-meter.

\bibliographystyle{naturemag}

\bibliography{main_ref.bib}